\documentstyle [12pt]{article}
\newcommand{\f}{\frac}
\newcommand{\br}{\bar{\rho}}
\newcommand{\nea}{\stackrel{\rho}{\nearrow}}
\newcommand{\sea}{\stackrel{\br}{\searrow}}
\newcommand{\p}{\partial}
\newcommand{\FR}{Fun(\rn_q^N)}
\newcommand{\DFR}{Dif\!f(\rn_q^N)}
\newcommand{\FS}{Fun(SO_q(N))}
\newcommand{\FE}{Fun(E_q^N)}
\newcommand{\FBE}{Fun(\bar E_q^N)}

\newcommand{\bp}{\bar{\partial}}
\newcommand{\s}{\star}
\newcommand{\e}{\vec{e}}
\newcommand{\y}{\vec{y}}
\newcommand{\w}{\vec{w}}
\newcommand{\1}{{\bf 1}}
\newcommand{\J}{\vec{j}}
\newcommand{\Pg}{\vec{\pi}}
\newcommand{\ot}{\otimes}
\newcommand{\La}{\Lambda}

\newcommand{\ve}{{\varepsilon}}
\newcommand{\ap}{\approx}
\newcommand{\bc}{\begin{center}}
\newcommand{\ec}{\end{center}}
\newcommand{\be}{\begin{equation}}
\newcommand{\ee}{\end{equation}}

\newcommand{\und}{\underline}

\newcommand{\uot}{\underline{\ot}}

\newcommand{\M}{{\bf L}}
\newcommand{\k}{{\bf k}}

\def\lcross{{>\!\!\!\triangleleft}}

\newcommand{\cn}{{\bf C}}
\newcommand{\rn}{{\rm\bf R}}
\newcommand{\zn}{{\bf Z}}
\newcommand{\nn}{{\rm\bf N}}
\newtheorem{prop}{Proposition}
\newtheorem{lemma}{Lemma}
\newtheorem{theorem}{Theorem}
\newtheorem{corollary}{Corollary}

\title{\huge{\bf The Euclidean Hopf Algebra $U_q(e^N)$ and its fundamental
Hilbert space representations }}

\author{{\bf Gaetano Fiore$^1$} \\ \\
{\it Sektion Physik der Universit\"at M\"unchen, Ls. Prof. Wess,} \\
{\it Theresienstrasse 37, D-80333 M\"unchen, Germany}\\
e-mail: fiore@lswes8.ls-wess.physik.uni-muenchen.de\\
Sissa 53/94/EP, LMU-TPW 94-11, hep-th 9407195\\
 To appear in J. Math. Phys.}
\date{November 1994}
\begin{document}

\baselineskip=23pt
\maketitle
\section*{\center Abstract}

We construct the Euclidean Hopf algebra $U_q(e^N)$ dual of
$Fun(\rn_q^N\lcross SO_{q^{-1}}(N))$ by realizing it
as a subalgebra of the differential
algebra $\DFR$ on the quantum Euclidean space $\rn_q^N$; in fact,
we extend our
previous realization \cite{fio4} of $U_{q^{-1}}(so(N))$ within $\DFR$
through the introduction of q-derivatives as generators of q-translations.
The fundamental Hilbert space representations of $U_q(e^N)$ turn
out to be of highest weight type and rather simple
`` lattice-regularized '' versions of the classical ones.
The vectors of a basis of the singlet (i.e. zero-spin) irrep can be
realized as
normalizable functions on $\rn_q^N$, going to distributions in the
limit $q\rightarrow 1$.

\vskip2truecm

\bc{$^{1}$ Alexander von Humboldt-Stiftung fellow}
\ec

\newpage

\section{Introduction}

{}~~~~
One of the most appealing fact explaining the present interest for quantum
groups \cite {dr}
is perhaps the idea that they can be used to generalize the ordinary notion
of space(time) symmetry. This generalization is tightly coupled to a
radical modification of the ordinary notion of space(time) itself.
{}From this viewpoint inhomogenous group symmetries such as Poincar\'e's and
the
Euclidean one yield physically relevant candidates for quantum group
generalizations;
Minkowski space $M^4$ and Euclidean
$\rn^N$ one are then the corresponding space(time) manifolds.

A major physical motivations for such generalizations is
the desire to discretize space(time) (or momentum space) in a `` wise ''
way for QFT regularization purposes. As known, standard
lattices used in regularizing QFT do not carry representations of
discretized versions (in the form of discrete subgroups) of the
associated inhomogenous groups; actually, the notion of a group is too tight
for this scope. For instance, the Euclidean
cubic lattice is invariant only under a discretized version of the translation
subgroup of the Euclidean group,
but not of the rotation one. On the contrary, the notion
of symmetry provided by quantum groups is broad enough
to allow the existence of lattices whose points are mapped into each other
under the action of inhomogeneous q-groups,
as we will explicitly see in the case of the Euclidean symmetry.

One way to develop the inhomogeneous quantum-group
program is to introduce a pair consisting of
a `` homogeneous '' (e.g. Lorentz's, resp. $SO(N)$) quantum group and the
associated quantum space \cite{frt,man} comodule algebra;
the quantum group coacts homogeneously on the quantum space. The inhomogeneous
quantum group \cite{schl} can be constructed by doing the braided
semi-direct product
\cite{maj3} of the homogeneous one with the corresponding quantum space
itself, thought as braid group \cite{maj2} of translations.
The latter is the proper generalization of the ordinary group of
translations.

This approach is probably closest to most physicists' way of thinking of
and dealing with space(time) and its inhomogeneous symmetry; it
turns out to be a rather viable one for q-deformations, too.
Here we adopt it in considering the $N$-dimensional ($N\ge 3$) Euclidean space
$\rn_q^N$ introduced in
\cite {frt} and its $\rn_q^N\lcross SO_q(N)$ symmetry $\cite{maj2,schl}$.

In absence of deformations, the generators of Euclidean Lie algebra $e^N$
can be realized as differential operators acting on $Fun(\rn^N)$,
more precisely the generators of $so(N)$-rotations can be realized as
the angular momentum components and the generators of translations as
 pure derivatives respectively.
In other words the algebra $Fun(\rn^N)$ of
functions on $\rn^N$ is the base space of (reducible) representations
of $so(N)$, of the abelian group $\rn^N$ of translations, and
of $e^N=\rn \lcross so(N)$.

Since the structure of $\FS$ is intimely related \cite{frt} to the
structures of $\FR$ and of the $SO_q(N)$-covariant differential calculus
\cite{car} which can be built on it,
it is interesting to ask whether analogues of these
facts occur in the q-deformed case; in proper language, whether
$Fun(\rn_q^N)$ can
be considered as a left (or right) module of the universal
enveloping algebra $U_q(so(N))$, the latter being realized as some subalgebra
$U^N_q$ of the algebra of differential operators $Dif\!f(\rn^N_q)$ on
$\rn^N_q$, and of some suitable q-deformed version of the abelian
algebra of infinitesimal translation; altogether, whether
$Fun(\rn_q^N)$ can be considered as a left (or right) module of some
suitable q-deformed version of the euclidean algebra of u.e.a. type.

In this work we give positive answers to these questions.
Actually, we $construct$ (chapter III)
a q-deformed version $U_q(e^N)$ of the Euclidean algebra of u.e.a. type by
requiring that the previous conditions are satisfied. Moreover,
we study the fundamental Hilbert space representations
of $U_q(e^N)$ (chapter IV). Chapter II provides the
reader with the necessary preliminaries, so that the whole paper
is basically self-contained. For more details see Ref. \cite{fio5}.

The construction
is based on our work in Ref. \cite{fio4}, of which this is the natural
development. There we realized the Hopf algebra
$U_{q^{-1}}(so(N))$ as a subalgebra $U_q^N$ of $\DFR$;
more precisely, we found $\DFR$-realizations of the
Drinfeld-Jimbo \cite{dr} generators of $U_{q^{-1}}(so(N))$ and showed how the
Hopf structure of $U_{q^{-1}}(so(N))$ can be derived in a natural way from
the simple commutation relations between these generators and the
coordinates $x^i$ of $\rn_q^N$.

Legitimated by these positive results, in the present work we extend
$U_q^N$ into a new (closed) algebra $\hat u_q(e^N)$
by adding to its Drinfeld-Jimbo generators
the q-derivatives $\p^i$ as generators of translations (section III.1).
To endow $\hat u_q(e^N)$ with a Hopf structure we proceed as in the case of
$U_q^N$; we realize that we need to
enlarge $\hat u_q(e^N)$ further by introducing
one more generator $\La\in Dif\!f(\rn_q^N)$, generating dilatations.
We call $U_q(e^N)$ the resulting Hopf algebra.

$U_q(e^N)$ coincides with the Euclidean Hopf algebra $B\lcross \tilde
U_q(so(N))$ of u.e.a type
previously constructed in ref. \cite{maj2} (using Hopf algebra duality
arguments).
Actually,  in section III.2 we see that $U_q(e^N)$
can be considered as the dual of the Hopf algebras $\FE,\FBE$, where
we $E_q^N,\bar E_q^N$ are the two versions of $\rn_q^N\lcross SO_q(N)$
(corresponding to the two possible braidings, see section II.2).   The real
structure
of $Dif\!f(\rn_q^N)$ induces (only at the algebra level)
a real structure for $U_q(e^N)$.
The $*$-structure however is not compatible with the Hopf one, due to
the presence of $\La$; this is the dual situation of what happens for
$\FE,\FBE$, where the $*$-structure is incompatible with the algebra.
\footnote{This is an already known problem \cite{wess}, common to many
approaches to inhomogeneous quantum groups.
We think that the way out should be seeked in a non-standard way to realize
Hilbert space tensor products of elementary systems into composite ones,
and hope to report progress on this point elsewhere.}
In section III.3 we define a new set of generators
of the q-Euclidean algebra, in view of the study of representations,
in section III.4 we give the casimirs of $\hat u_q(e^N)$.

The whole construction is a very economic one, since it relies on the
use of only $2N$ objects $\{x^i,\p_j\}$
(the coordinates and derivatives, i.e. the generators of $Dif\!f(\rn_q^N)$)
with already fixed commutation, derivation and $*$
relations, giving all
the generators of $U_q(e^N)$, the corresponding algebra, Hopf algebra and
$*$-algebra structures, together with a bunch of representations
(the ones contained
in the reducible representation $\FR$, i.e. having
zero `` spin '').  Actually, the algebra relations of $U_q(e^N)$ follow
from the algebra relations of $\DFR$, and $U_q(e^N)$ gets a bialgebra
by deriving the natural coalgebra structure associated to it when we think
of its elements as differential operators on $\FR$;
the compatibility of the coalgebra with the algebra
follows from the associativity of
$\DFR$ and $\FR$; finally, the antipode is found by consistency.
Once the structure of $U_q(e^N)$ is determined in this way,
we abstractly postulate it, so as to allow for all representations, i.e.
representations with
arbitrary $U_q(so(N))$ highest weight.

$U_q(e^N)$ is in many aspects the Euclidean analogue
of the q-deformed Poincare' Hopf algebra of Ref. \cite{wess}.
In both cases the inhomogeneous Hopf algebra contains the
homogeneous one as a Hopf subalgebra which can be obtained by setting
$p^i=0,\La=1$. In fact, all the commutation relations in $U_q(e^N)$
are homogeneous in $p$, contrary to what happens for inhomogenous
Hopf algebras obtained through contractions \cite{cel,luk1,luk2}.

Chapter IV is devoted to the study of fundamental (i.e. irreducible
one-particle)
representations of $U_q(e^N)$;
since we are interested in potential applications to
quantum physics, we look only for Hilbert space ones. We will call them
`` irreps '' in the sequel.
Only the *-algebraic structure $u_q(e^N)$ of $U_q(e^N)$
is involved (the Hopf structure
is irrelevant at level of irreducible representations), so no problem
of compatibility of $*$ with the coalgebra arises.

In section IV.1 we do an abstract study. We choose a Cartan subalgebra
(i.e. a complete set of commuting observables) consisting basically of
two parts, $[\f {N+1}2]$ squared momentum components and $[\f N2]$
angular momentum components. In subsection IV.1.1, IV.1.2 we do their spectral
analysis separatly;
this is possible by means of the $L$ generators introduced in
section III.3. The points of the spectra make up a q-lattice.
One important fact is that the irreps turn out to be
of highest weight type, and they can
be obtained from tensor products of a single one (the singlet, i.e.
the one describing a particle with zero $U_q^N=U_{q^{-1}}(so(N))$
highest weight) with some representation
of $U_{q^{-1}}(so(N))$. The spectra of all
observables are discrete, in particular the spectra of squared momentum
components, as expected. The corresponding eigenvectors
are normalizable and make up an orthogonal basis of the Hilbert space
of each irrep. In other words we have `` lattice-regularized '' the irrep,
in the sense mentioned before.
In subsection IV.1.3 we concentrate on the structure of the
singlet representation. Moding out singular vectors amounts to
strenghtening the Serre relations of the $L$-roots;
a cumbersome `` kinematical parity asymmetry '' appears
in the structure of the spectra of the angular momentum observables,
which is an anusual feature for a `` lattice '' theory \cite{nie}; of course,
it disappears in the limit $q\rightarrow 1$.

In section IV.2 we find a nonstandard
configuration-space realization of the abstract singlet representation,
actually based on a $pair$ (the unbarred and the barred) realizations of the
abstract vectors as functions on $\rn_q^N$; the two realizations
are put together in the definition of the scalar product as a
$\rn_q^N$-space integral.
For instance, in the $N=2n+1$ the highest weight vector
is represented by a q-plane wave directed in the $x^0$-direction.
A suggestive final comment to sections IV.1, IV.2 singles out that
ultimately the
$x^i$-coordinate description of the space of physical states
(heuristically,
the Euclidean space itself) arises naturally from the existence of its
Hilbert-space structure.

In section IV.3 we briefly analyze the
limit $q\rightarrow 1$ of the representations.

{}~

We can think of the irreps studied in chapter I as
describing the (time-independent) dynamics of
a free nonrelativistic particle with arbitrary `` generalized ''
$U_{q^{-1}}(so(N))$-spin on $\rn_q^N$.
The subalgebra $\hat u_q(e^N)=u_q(e^N)/\{\La=1\}$
can be considered as the quantum group symmetry of the hamiltonian
\be
H:=\f{(p\cdot p)}{2M}.
\ee
of the system; therefore
all states with a given energy should be obtained from each other
by the action of $\hat u_q(e^N)$, as in the classical case,
different eigenspaces of the energy should be obtained from each other by the
action of the dilatation operators $\La^{\pm 1}$. More interestingly
the study of chapter IV can be considered as a warm-up before the
construction of positive mass q-Wick-rotated
versions of the representations \cite{piwe} of the q-Poincar\'e algebra
of Ref. \cite{wess,mey}, in view of a q-Euclidean formulation of QFT.
In fact, the q-deformed version of Wick-rotation in $3+1$ dimensions
\cite{majw} rotates our subalgebra $\hat u_q(e^4)$ into
the $q$-deformed Poincare'algebra of that Ref.
For such (socalled virtual) representations one has to impose a
condition equivalent to
the positivity of the energy in Minkowski spacetime\cite{froh}.
 We hope to report on this point elsewhere.

{}~

 There are some alternative approaches to
inhomogeneous quantum groups.
A different (and chronologically preceding) approach is based on
contractions of homogeous quantum groups of higher rank, see for instance
Ref. \cite{cel,luk1,luk2}; another one based on projections of
bicovariant differential calculi on homogeous quantum groups of higher rank
has been recently proposed in Ref. \cite{cast}

Some notational remarks are necessary before the beginning.
We will assume that $q$ is generic; in chapter IV we have to consider
$q\in {\bf R}^+$, and we will assume that $0<q\le 1$;
the case $q>1$ can be treated in an
analogous way, and in doing so one essentially will interchange the roles of
annihilation and creation operators.
We set $h=h(N)=\cases{0~~~if~~N=2n+1\cr 1~~~ if~~N=2n\cr}$ to allow
a compact way of writing relations valid both for even and odd $N$.
In our notation a space index $i$ takes the values
$i=-n,-n+1,...,-1,0,1,...n$ if
$N=2n+1$, and $i=-n,-n+1,...,-1,1,...n$ if $N=2n$.
When $N=2n$ there is a complete symmetry of $Dif\!f(\rn_q^N)$ under
the exchange $x^{-1},\p^{-1}\leftrightarrow x^1,\p^1$,
and this implies a complete invariance of the validity of all
the results of this chapter and of the following ones
under the exchange of indices $-1\leftrightarrow 1$, so that we will
normally omit writing down explicitly the results that can be obtaind
by such an exchange. We will often use the shorthand notation
$[A,B]_a:=AB-aBA$ ($\Rightarrow [\cdot,\cdot]_1=[\cdot,\cdot]$).
Finally, in our conventions $\nn:=\{0,1,2,...\}$.

{}~~

\tableofcontents

\newpage
\section{Preliminaries}

{}~~~~
In this chapter we recollect some
basic definitions and relations characterizing:
\begin{itemize}
\item 1) the algebra $Fun({\bf R}_q^N)$
($O_q^N({\bf C})$ in the notation of \cite{frt}) of
functions on the quantum euclidean space ${\bf R}_q^N$, $N\ge 3$, and
the algebra $Dif\!f({\bf R}_q^N)$ of
differential operators on ${\bf R}_q^N$;

\item 2) the quantum group $\FS$ \cite{frt} and its
inhomogeneous extension \cite{schl} $\FE$, $E_q^N:={\bf R}_q^N\lcross SO_q(N)$;

\item 3) the exterior algebra on $\rn_q^N$ and the q-epsilon tensor;

\item 4) the action of the Hopf
algebra $U_q^N\ap U_{q^{-1}}(so(N))$ on $\FR$ as determined
in Ref. \cite{fio4}.

\end{itemize}

For further details we refer the reader to Ref.
\cite{frt,car,og,ogzu,fio4}.

\subsection{$\FR,\DFR$}

{}~~~ $\hat R_q:=||\hat R^{ij}_{hk}||$ is the braid matrix
for the quantum group $SO_q(N)$ (it can be found for instance in Ref.
\cite{frt,car});
$\hat R$ is symmetric: $\hat R^t=\hat R$.

{}~~~The q-deformed metric matrix $C:=||C_{ij}||$ is explicitly given by
\be
C_{ij}:=q^{-\rho_i}\delta_{i,-j},
\ee
where
\be
(\rho_i):=\cases {
(n-\f{1}{ 2},n-\f{3}{ 2},...,\f{1}{ 2},0,-\f{1}{ 2}...,\f{1}{ 2}-n)
{}~~~~~~~~~~if~N=2n+1 \cr
(n-1,n-2,...,0,0,...,1-n)~~~~~~~~~~~~~~~if~N=2n. \cr}
\ee
Notice that $N=2-2\rho_n$ both for even and odd $N$.
$C$ is not symmetric and coincides with its inverse: $C^{-1}=C$.
Indices are raised
and lowered through the metric matrix $C$,  for instance
\be
a_i=C_{ij}a^j,~~~~~~a^i=C^{ij}a_j,
\ee

Both $C$ and $\hat R$
depend on $q$ and are real for
$q\in {\bf R}$. $\hat R$ admits the very useful decomposition
\be
\hat R_q = q {\cal P}_S - q^{-1} {\cal P}_A +q^{1-N}{\cal P}_1.
\ee
${\cal P}_S,{\cal P}_A,{\cal P}_1$ are the projection operators
onto the three eigenspaces of $\hat R$ (the latter have respectively dimensions
$\f{{N(N+1)}}{ 2}-1,\f{{N(N-1)}}{ 2},1$): they project the tensor
product $x\otimes x$ of the fundamental corepresentation $x$ of $SO_q(N)$
into the corresponding irreducible corepresentations (the symmetric modulo
trace, antisymmetric and trace, namely the q-deformed
versions of the corresponding ones of $SO(N)$). The projector ${\cal P}_1$ is
related to the metric matrix $C$ by ${\cal P}_{1~hk}^{~~ij}=\f{C^{ij}C_{hk}}{
Q_N}$; the factor $Q_N$ is defined by $Q_N:=C^{ij}C_{ij}$.
$\hat R^{\pm 1},C$ satisfy the relations
\be
[f(\hat R), P\cdot (C\otimes C)]=0~~~~~~~~~
f(\hat R_{12})\hat R^{\pm 1}_{23}\hat R^{\pm 1}_{12}=
\hat R^{\pm 1}_{23}\hat R^{\pm 1}_{12}f(\hat R_{23})
\ee
($P$ is the permutator: $P^{ij}_{hk}:=\delta^i_k\delta^j_h$ and $f$ is any
rational function);
in particular this holds for $f(\hat R)=\hat R^{\pm 1},{\cal P}_A,{\cal P}_S,
{\cal P}_1$.

{}~~~Let us recall that the unital
algebra $Dif\!f({\bf R}_q^N)$ of differential operators on the real quantum
Euclidean plane ${\bf R}_q^N$ is essentially
defined as the space of formal series in the
(ordered)
powers of the $\{x^i\},\{\p_i\}$ variables, modulo the commutation relations
\be
{\cal P}_{A~hk}^{~~ij}x^h x^k =0     ~~~~~~~~~~~~~
{\cal P}_{A~hk}^{~~ij}\p^h \p^k =0
\ee
and the derivation relations
\be
\p_i x^j = \delta^j_i+q\hat R^{jh}_{ik} x^k\p_h
\ee
(actually we will enlarge it a bit more by allowing also arbitrary integer
powers of the dilatation operator $\La$, which is defined below).
The subalgebra $Fun({\bf R}_q^N)$ of `` functions '' on ${\bf R}_q^N$ is
generated by $\{x^i\}$ only. Below we will give the explicit form of these
relations.

For any function $f(x) \in Fun({\bf R}_q^N)$
$\p_i f$ can be expressed in the form
\be
\p_i f= f_i + f_i^j\p_j,~~~~~~~~~~f_i,f_i^j\in Fun({\bf R}_q^N)
\ee
(with $f_i,f^i_j$ uniquely determined)
upon using the derivation relations () to move step by step the derivatives
to the right of each $x^l$ variable of each term of
the power expansion of $f$, as far as the extreme right.
We denote $f_i$ by $\p_i f|$. This defines the action of $\p_i$ as a
differential operator
$\p_i:f\in Fun({\bf R}_q^N)\rightarrow \p_if|\in Fun({\bf R}_q^N)$: we will
say that $\p_if|$ is the `` evaluation '' of $\p_i$ on $f$. For instance:
\be
\p_i {\bf 1}|=0,~~~~~~~~\p^ix^j|=C^{ij},~~~~~~~~\p^ix^jx^k|=C^{ij}x^k+
q\hat R^{-1~ij}_{~~~hl}x^hC^{lk}
\ee
By its very definition, $\p_i$ satisfies the generalized Leibnitz rule:
\be
\p_i(fg)|=\p_i f|g+({\cal O}_i^jf)\p_jg|,~~~~~~~~~f,g\in Fun({\bf R_q^N}),
\ee
$({\cal O}_i^jf)=f_i^j$. In
a similar way one can define \cite{fio4} the evaluation of differential
operators
corresponding to the angular momentum components. One of the results of
this paper will be that we will be able to write explicitly the linear
operators
${\cal O}_i^j$ as differential operators,
${\cal O}_i^j\in \DFR$.

{}~~~If $q\in {\bf R}$ one can introduce an antilinear involutive
antihomomorphism
$*$:
\be
*^2=id~~~~~~~~~~~~~~~~~~~~~~~(AB)^*=B^*A^*
\ee
on $Dif\!f({\bf R}_q^N)$.  On the basic variables $x^i$  $*$ is
defined by
\be
(x^i)^*=x^jC_{ji}
\ee
whereas the complex conjugates of the derivatives $\p^i$ are not
combinations of the derivatives themselves. It is useful to introduce
barred derivatives $\bar \p^i$ through
\be
(\p^i)^*=-q^{-N}\bar \p^j C_{ji}.
\ee
They satisfy relation $(7)$ and the analogue of (8) with
$q,\hat R$ replaced by $q^{-1},\hat R^{-1}$. These $\bar\p$ derivatives can
be expressed as functions of $x,\p$ \cite{og}, see formula (22).

Let us define
\be
(a\cdot b)_j:=\sum\limits_{l=1}^j a^{-l}b_{-l}+\cases{\f{a^0b_0}{1+q^{-1}}
{}~~~~if~~N=2n+1\cr 0~~~~if~~N=2n, \cr}
{}~~~~~~~~~~~~ and~~~0< j\le n
\ee
(when this causes no confusion we will also use the notation
$a\cdot b:=(a\cdot b)_n$).

Relations (7), (8) defining $Dif\!f({\bf R}_q^N)$ amount respectively to
\be
\eta^i\eta^j=q\eta^j\eta^i,~~~~~~i<j,
{}~~~~~~~~~~~~~~~~
\sum\limits_{l=-j}^j\eta^l\eta_l=(1+q^{-2\rho_j})(\eta\cdot \eta)_j
{}~~~~~~~j=1,2,...,n;~~~~~~~~~~\eta^i=x^i,\p^i,
\ee
and
\be
\cases{
\p_kx^j=qx^j\p_k-(q^2-1)q^{-\rho_j-\rho_k}x^{-k}\p_{-j},~~~
{}~~~~~~~~~~~~~~~~~~j<-k,j\neq k \cr
\p_kx^j=qx^j\p_k
{}~~~~~~~~~~~~~~~~~~~~~~~~~~~~~~~~~~~~~~~~~~~~~~~~~~~~~~~j>-k,j\neq k \cr
\p_{-k}x^{k}=x^k\p_{-k},~~~~~~~
{}~~~~~~~~~~~~~~~~~~~~~~~~~~~~~~~~~~~~~~~~~~~~~~~~~~~~~~~~~k\neq 0  \cr
\p_ix^i=1+q^2x^i\p_i+(q^2-1)\sum\limits_{j>i}x^j\p_j,
{}~~~~~~~~~~~~~~~~~~~~~~~~~~~~~~~~~~~i>0 \cr
\p_ix^i=1+q^2x^i\p_i+(q^2-1)\sum\limits_{j>i}x^j\p_j
-q^{-2\rho_i}(q^2-1)x^{-i}\p_{-i},
{}~~~~~~~i<0 \cr
\p_0x^0=1+qx^0\p_0+(q^2-1)\sum\limits_{j>0}x^j\p_j,
{}~~~~~~~~~~~~~~~~~~~~~~~(only~for~N~odd). \cr}
\ee
Here are some useful formulae for the sequel (sum over $l$ is understood):
\be
\p^i (x\cdot x)_n=q^{2\rho_n}x^i+q^2(x\cdot x)_n\p^i~~~~~~~
(\p\cdot \p)_n x^i=q^{2\rho_n}\p^i+q^2x^i(\p\cdot\p)_n
\ee
\be
(x^l\p_l) x^i=x^i+q^2x^i x^l\p_l +(1-q^2)(x\cdot x)_n \p^i~~~~~~~
\p^i (x^l\p_l)=\p^i+q^2(x^l\p_l) \p^i+(1-q^2)x^i(\p\cdot \p)_n.
\ee

The dilatation operator $\La$ is given by
\be
\La^2:=1 +(q^2-1)x^i\p_i+q^{N-2}(q^2-1)^2(x\cdot x)(\p\cdot\p)
\ee
(note the change of convention $\La\rightarrow \La^2$ w.r.t. Ref.
\cite{fio4,og}); it fulfils the relations
\be
\La x^i=qx^i\La,~~~~~~~~~~\La\p^i=q^{-1}\p^i\La.
\ee
Then one can prove \cite{og} that
\be
\bar \p^k=\La^{-2}[\p^k+q^{N-2}(q^2-1)x^k(\p\cdot\p)]
\ee
The operator $B$ is defined by
\be
B:=\La^{-1}[1+q^{N-2}(q^2-1)(x\cdot\p)]=:\La^{-1}{\cal B}
\ee
and satisfies the relations
\be
[B,(x\cdot x)]=0=[B,(\p\cdot\p)].
\ee
Under complex conjugation
\be
\La^*=q^{-N}\La^{-1}~~~~~~~~~~~~~B^*=B.
\ee

{}~~~By definition a scalar $I(x,\p)\in Dif\!f({\bf R}_q^N)$ transforms
trivially
under the coaction $\phi_L$ (defined in subsection II.2)
associated to the quantum group of symmetry
$SO_q(N,{\bf R})$ \cite{frt}: $\phi_L(I)={\bf 1}\ot I$.
In Ref. \cite{fio2} we showed that
any scalar $I(x,\p)\in Dif\!f({\bf R}_q^N)$  can be written as a function
of $x\cdot x,\p\cdot \p$. Using relations (20),(23) it is straightforward
to verify that the product $(x\cdot x)(\p\cdot\p)$ can be written as a
linear combination of $\La^2,B\La,1$. Therefore any scalar $I$ with natural
dimension zero can also be written as a function $I=I(B,\La)$ only.

\subsection{The Euclidean quantum groups $E_q^N$, $\bar E_q^N$}

$E_q^N,\bar E_q^N$ can be introduced as braided semidirect products
$\rn_q^N\lcross SO_q(N)$ \cite{maj2} with the two possible covariant
braidings; in this way the braided Hopf algebra $\rn_q^N$ is
embedded into an ordinary one, a process which was called `` bosonization ''
\cite{maj2}. The result is the one which was found in Ref \cite{schl} and
which we briefly review here.

The Euclidean Hopf algebra $\FE:=Fun(\rn_q^N\lcross SO_q(N))$
as an algebra is unital and is generated by
elements $w^{\pm 1},T^i_j,y^i$ satisfying relations
\be
\hat R^{ij}_{hk} T^h_lT^k_m=T^i_hT^j_k\hat R^{hk}_{lm},
{}~~~~~~~~~~~~~~
T^i_jC^{jl}T^k_l={\bf 1}_{E_q^N}C^{ik}=T_i^jC_{jl}T_k^l,
\ee
\be
{\cal P}_{A~hk}^{~~ij}y^h y^k =0,
\ee
and cross relations
\be
wy^i=q^{-1}y^iw,~~~~~~~~~~[w,T^i_j]=0,~~~~~~~~~~~y^iT^j_h=\hat R^{-1~ij}
_{~~~lm}T^l_hy^m;
\ee
${\bf 1}_{E_q^N}$ denotes the unit of the algebra.
The coproduct, counit and antipode
 $\phi^E,\ve^E,S^E$ on the generators $w^{\pm 1},T^i_j$ are defined by
\be
\phi^E(T^i_j)=T^i_h\ot T^h_j~~~~~~~~~~~~~\ve(T^i_j)=1~~~~~~~~~~~~~~
S^E(T^i_j)=C^{il}T^m_lC_{mj},
\ee
\be
\phi^E(w)=w\ot w~~~~~~~~~~~~~~~\ve(w)=1~~~~~~~~~~~~~S^E(w)=w^{-1}
\ee
\be
\phi^E(y^i)=w^{-1}T^i_j\ot y^j+y^i\ot {\bf 1}~~~~~~~~~~~
\ve^E(y^i)=0~~~~~~~~~~~S^E(y^i)=-wS(T^i_j)y^j=-w(CT^tC)^i_jy^j.
\ee
$\phi^E,\ve^E,S^E$ are extended as algebra (anti)homomorphisms as usual.
$\FS$ is the Hopf subalgebra generated by $T^i_j$ only.

Similarly we define the Euclidean Hopf algebra
$\FBE:=Fun(\rn_q^N\lcross' SO_q(N))$ (corresponding
to the second choice of the braiding)
by replacing $q,\hat R,w\rightarrow q^{-1},\hat R^{-1}, w^{-1}$
in formulae (28)-(31); the corresponding generators of translations
will be denoted by $\bar y$ (instead of $y$). Note a change of notation
w.r.t to ref. \cite{schl}: our $y^i,T^j_k$ are $\bar x^i,M^j_k$ in the notation
of that reference.

When $q\in \rn$ we can define the complex conjugation $*$ as
an antilinear involutive antihomomorphism mapping
$\FE\leftrightarrow \FBE$ by defining it on the generators through the
formulae
\be
(T^i_j)^*=C^{li}T^i_mC_{jm}~~~~~~~~~~~~~\bar y^i=(y^j)^*C^{ji}~~~~~~~~~~~~
w^*=w^{-1}.
\ee
One can immediately verify that we cannot
impose the reality condition $y^i=(y^j)^*C^{ji}$ which we adopted for
the quantum space $\rn_q^N$ (formula (13)), otherwise we would
spoil the algebra relations $(28)_3$.

Finally $\FR$ can be equipped with a left (or right, as well)
$E_q^N$- and $\bar E_q^N$-comodule structure introducing respectively
the left coactions defined on the basic generators by
\be
\phi^E_L(x^i):=w^{-1}T^i_j\ot x^j+ y^j\ot{\bf 1}~~~~~~~~or~~~~~~~~
\phi^E_L(x^i):=wT^i_j\ot x^j+ \bar y^j\ot{\bf 1},
\ee
and is extended as an algebra homomorphism;
the $SO_q(N)$ left coaction $\phi_L$ can be obtained from $\phi^E_L$
by setting $w=1,~y=0$ in the previous formula.

\subsection{q-Epsilon tensor and Hodge duality}

The exterior algebra $\bigwedge({\bf R}_q^N)$ over ${\bf R}_q^N$ is defined
as the algebra generated by `` 1-forms '' $\xi^i$ satisfying the
relations
\be
{\cal P}_S\xi\xi=0={\cal P}_1\xi\xi;
\ee
it is equipped with the same comodule structure as $Fun({\bf R}_q^N)$.
In Ref. \cite{fio3} we showed that, as in the classical case,
$\bigwedge({\bf R}_q^N)$ admits only subspaces of forms of degree $m\le N$, and
that there exists a
unique $N$-form (the `` volume '' form $dV$), which transforms
as a scalar (since it is mapped by the coaction to itself times the
$q$-determinant, and the latter is essentially equal to one). The existence
of such a $N$-form implies the existence of a $q$-deformed
completely antisymmetric tensor with $N$ indices
$\ve^{i_1i_2...i_N}$
with the same fundamental properties as in the classical case, by
\be
\xi^{i_1}\xi^{i_2}...\xi^{i_N}=dV\ve^{i_1i_2...i_N}
{}~~~~~~~~~~~~~~~~dV:=\xi^{-n}\xi^{1-n}...\xi^n
\ee

We give the explicit expression for the tensor $\ve_q$
in the case $N=3,4$:
\be
\begin{array}{|c|c|c|c|}
\hline
\ve_q^{-101}=1 & \ve_q^{-110}=-q & \ve_q^{0-11}=-q & \ve_q^{01-1}=q \\
\hline
\ve_q^{10-1}=-q^2 & \ve_q^{1-10}=q &
\ve_q^{000}=-q(q^{1\over 2}-q^{-1\over 2}) & \ve_q^{ijk}=0~~~otherwise \\
\hline
\end{array}
\ee
\be
\begin{array}{|c|c|c|c|}
\hline
\ve_q^{-2-112}=1 & \ve_q^{-21-12}=-1 &\ve_q^{-2-121}=-q &\ve_q^{-212-1}=q\\
\hline
\ve_q^{-22-11}=q^2 & \ve_q^{-221-1}=-q^2 & \ve_q^{-1-212}=-q &
\ve_q^{-11-22}=q^2 \\
\hline
\ve_q^{-1-221}=q^2 & \ve_q^{-12-21}=-q^2 & \ve_q^{-121-2}=q^3 &
\ve_q^{-112-2}=-q^2 \\
\hline
\ve_q^{1-1-22}=-q^2 & \ve_q^{1-2-12}=q & \ve_q^{1-12-2}=q^3 &
\ve_q^{12-1-2}=-q^3 \\
\hline
\ve_q^{12-2-1}=q^2 & \ve_q^{1-22-1}=-q^2 & \ve_q^{2-2-11}=-q^2 &
\ve_q^{2-1-21}=q^3 \\
\hline
\ve_q^{21-2-1}=-q^3 & \ve_q^{2-21-1}=q^2 & \ve_q^{2-11-2}=-q^4 &
\ve_q^{21-1-2}=q^4 \\
\hline
\ve_q^{-11-11}=q(q^2-1) & \ve_q^{1-11-1}=-q(q^2-1) &
\multicolumn{2}{|c|}{\ve_q^{ijkl}=0~~~otherwise} \\
\hline
\end{array}
\ee
Exactly as in the classical case we can define a Hodge duality operation
$^{\s}:\wedge({\bf R}_q^N)\rightarrow \wedge({\bf R}_q^N)$ by
\be
^{\s}\xi^{i_1}...\xi^{i_m}:=A_m\ve^{i_1i_2...i_N}\xi_{i_{m+1}}...\xi^{i_N};
\ee
the normalization constant $A_m$ is such that $(^{\s})^2=\pm 1$.
It ~is ~easy~ to~ verify that
$(\xi^{i_1}...\xi^{i_m}) ^{\s}(\xi^{i_1}...\xi^{i_m})$ transforms
under the coaction as a scalar.

\subsection{The $*$-Hopf algebra $U_q^N\ap U_{q^{-1}}(so(N))$}

\begin{theorem}(\cite{fio4})
The Hopf algebra of u.e.a. type $U_{q^{-1}}(so(N))$ can be realized as
`` the algebra of the angular momentum on ${\bf R}_q^N$ '', i.e. the subalgebra
$U_q^N$ of $Dif\!f({\bf R}_q^N)$ whose elements commute with any
scalar $I(x,\p)\in Dif\!f({\bf R}_q^N)$.
\end{theorem}

A Poincar\'e-Birkhoff-Witt basis of $U_q^N$ is provided by the generators
$\{B,l^{ij}\}$, where $B$ was defined in formula (23) and
\be
l^{ij}:=\La^{-1}{\cal P}_{A~hk}^{~~~ij}x^h\p^k=-q^{-2}
\La^{-1}{\cal P}_{A~hk}^{~~~ij}\p^hx^k
\ee
(note the change of convention $L\rightarrow l$ w.r.t. Ref. \cite{fio4}).
Of course the $l^{ij}$ are not all algebraically independent and have rather
involved commutation relations. $B$ is a function \cite{fio4}
of the quadratic casimir
$l^{ij}l_{ji}$ only.

There exist closed commutation relations between the generators $l^{ij},B$
and $v^i$ ($v^i=x^i,\p^i,\bp^i$ etc.). We derive here only the commutation
relations with $B$:

\begin{prop}\cite{fio5}
\be
v^iB=q^{-1}\f{q^N+1}{1+q^{N-2}}Bv^i+\f{(q^{-1}-q)(q^4-1)}
{(1+q^{N-2})(1+q^{4-N})}l^{ib}v_b
\ee
\end{prop}
$Proof$. Using formula
(5) the definitions (15),(39), and equations (23) it is easy to prove
the following relation:
\be
x^i(\p\cdot \p)=\f{x^l\p_l}{1+q^{N-2}}\p^i+ \f{(q^2+1)q^{2-N}}{1+q^{4-N}}
{\cal L}^{ij}\p_j.
\ee
Then,
$$
{\cal B}\p^i\stackrel{(19)}{=}q^{-2}\p^i{\cal B}+\f{1-q^{-2}}{1+q^{2-N}}
\left[q^{2-N}\p^i+(q^2-1)x^i(\p\cdot\p)\right]
$$
\be
\stackrel{(41)}{=}q^{-2}\p^i{\cal B}+\f{1-q^{-2}}{1+q^{2-N}}
\left[q^{2-N}{\cal B}\p^i+
\f{(1-q^{-2})(q^4-1)}{(1+q^{N-2})(1+q^{4-N})} {\cal L}^{ij}\p_j\right].
\ee
We find relation (40) if we
collect terms in ${\cal B}\p^i$ together, multiply both sides of this identity
by $q^2\La^{-1}$, use relation (21) and the definition $B={\cal B}\La^{-1}$.
$\diamondsuit$

A more manageable
Poincar\'e-Birkhoff-Witt basis of $U_q^N$
is provided by the set $\{\M^{ij},\k^l\}$ ($i< j,\neq -j;~n\ge l>1$),
which was
constructed in Ref. \cite{fio4}; $\k^i$'s generate a Cartan subalgebra.
In particular the elements $\M^{-i,i+1},\M^{-i-1,i},\k^{i-1}(\k^i)^{-1}$
$i=h,h+1,...,n$ are
`` Chevalley generators '' of $U_q^N$, coinciding with the Drinfeld-Jimbo
generators of $U_q^N$ up to rescaling of the roots $\M$ by suitable
functions of $\k^i$.
All the other roots $\M^{ij}$ can be constructed starting from
them as follows:
\be
[\M^{-jl},\M^{-lk}]_q=q^{\rho_l}\M^{-j,k}~~~~~~~~~~[\M^{-kl},\M^{-l,j}]
_q=q^{\rho_l+1}\M^{-k,j},~~~~~~~~~~~~~~n\ge k>l>j\ge -h(N)
\ee
\be
[\M^{l-1,k},\M^{1-l,l}]_{q^{-1}}=q^{\rho_l-1}\M^{lk}~~~~~~~~~~
[\M^{-l,l-1},\M^{-k,1-l}]_{q^{-1}}=q^{\rho_l}\M^{-k,-l}~~~~~~~~~~~~~~~~~~~~~
2\le l<k\le n
\ee
\be
[\M^{0k},\M^{01}]=q^{-1}\M^{1k}~~~~~~~~~~[\M^{-10},\M^{-k0}]=\M^{-k,-1}
{}~~~~~~~~~~~~~~~~~~~~~~~1<k\le n~~~if~~N=2n+1.
\ee
Then the commutation relations satisfied by the Chevalley generators can be
summarized in the following way.
Commutation relations between the generators of the Cartan subalgebra and
the simple roots:
\be
[\k^i,\M^{\pm (1-k),\pm k}]_a=0~~~~~~a=\cases{q^{\pm 2}~~if~i=k\le n \cr
q^{\mp 2}~~~if~~i=k-1 \cr 1~~~otherwise}~~~~~~~~~~~~~~~~~~
[\k^i,\k^j]=0;
\ee
commutation relations between positive and negative simple roots:
\be
[\M^{1-m,m},\M^{-k,k-1}]_a=0~~~~~~~~~~~~~~~a=\cases{q^{-1}~~~~m\pm 1=k \cr
1~~~~if~~k\neq m,m\pm 1\cr}~~~~m,k\ge h(N)+1,
\ee
\be
[\M^{12},\M^{-2,1}]=0~~~~~~~~~~[\M^{-1,2},\M^{-2,-1}]=0~~~~~~~~~~~
{}~~~~~~~~~~if~~~~~N=2n,
\ee
\be
\cases{[\M^{1-m,m},\M^{-m,m-1}]_{q^2}=q^{1+2\rho_m}\f{1-\k^{m-1}(\k^m)^{-1}}
{q-q^{-1}}~~~~~~~~2\le m\le n\cr
[\M^{01},\M^{-1,0}]_q=q^{-\f 12}\f{1-(\k^1)^{-1}}{q-q^{-1}}~~~~~~~~
if~~N=2n+1;\cr}
\ee
Serre relations:
\be
[\M^{1-m,m},\M^{1-k,k}]=0~~~~~~~~~[\M^{-m,m-1},\M^{-k,k-1}]=0~~~~~~~~~~~~~~~
m,k>0,~~~~ |m-k|>1
\ee
\be
[\M^{1+j-m,m-j} ,\M^{2-m,m}]_a=0=[\M^{-m,m-2},\M^{j-m,m-j-1}]_a
{}~~~~~~~~~~~~~~a=\cases{q~~~if~~j=0\cr q^{-1}~~~
if~~j=1\cr}~~~~~~~~m\ge 3
\ee
\be
\cases{[\M^{01},\M^{12}]_{q^{-1}}=0 \cr [\M^{-1,2},\M^{02}]_q=0\cr}
{}~~~~~~~~~
\cases{[\M^{-2,-1},\M^{-1,0}]_{q^{-1}}=0\cr [\M^{-2,0},\M^{-2,1}]_q=0\cr}
{}~~~~~~~~~~~~~~~~~~~~~~if~~N=2n+1.
\ee

When $q\in {\bf R}$
the complex conjugation (13),(14) acts on the Chevalley generators in the
following way:
\be
(\k^i)^*=\k^i,~~~~~~~~~(\M^{1-k,k})^*=q^{-2}\M^{-k,k-1}~~~~~~k\ge 2,
{}~~~~~~~~~~~~\cases{(\M^{01})^*=q^{-\f 32}\M^{-10}~~~~~if~~N\!=\!2n+1\cr
(\M^{12})^*=q^{-2}\M^{-2,-1}~~~~~if~~N\!=\!2n\cr}
\ee

There exist closed commutation relations between the generators
of $U_q^N$ and the coordinates $x^i$. Let $m\ge h(N)+1$. Then
\be
[\k^h,x^i]_{a_{h,i}}=0,~~~~~~~h=1,2,...,n;
\ee
\be
[\M^{01},x^0]=-q^{-1}x^1~~~~~~~~~~~~~[\M^{-1,0}x^0]=x^{-1}~~~~~~~~~~~~~~~~~~
{}~~~~~~~if~~N=2n+1,
\ee
and in all the remaining cases
\be
[\M^{1-m,m},x^i]_{b_{m,i}}=q^{\rho_m}(\delta^i_{-m}-\delta^i_{m-1})x^{i+1}
{}~~~~~~~~~~~[\M^{-m,m-1},x^i]_{b_{m,i}}=q^{\rho_m}(\delta^i_{1-m}-\delta^i_m)
x^{i-1},
\ee
where
\be
a_{m,i}:=q^{2(\delta^i_m-\delta^i_{-m})},~~~~~~~~~~~b_{m,i}:=
(a_{m-1,i})^{\f 12}(a_{m,i})^{-\f 12}.
\ee

The commutation relations of $\k^i,\M^{ij}$'s with $\p_i,\bar \p_i$ are the
same, since $\p_i\propto [\p\cdot\p,x_i]_{q^2}$,
$\bar\p_i\propto [\bar\p\cdot\bar\p,x_i]_{q^{-2}}$ and the $\k,\M$'s commute
with
scalars.
Considering $\M^{ij}$'s, $\k^i$'s as differential operators on
$Fun({\bf R}_q^N)$ (see formula (11)) one gets the Hopf algebra structure of
$U_{q^{-1}}(so(N))$ in a natural way \cite{fio4}.
In particular the Leibnitz rule of
these differential operators determines its coassociative coproduct
$\phi:U_q^N\rightarrow U_q^N\bigotimes U_q^N$,
which on the generators $\k^i,\M^{ij}$ takes the form
\be
\phi(\k^i)=\k^i\bigotimes \k^i
\ee
\be
\cases{\phi(\M^{1-m,m})=\M^{1-m,m}\bigotimes
{\bf 1'}+ (\k^{m-1}(k^m)^{-1})^{\f 12}\bigotimes \M^{1-m,m} \cr
\phi(\M^{-m,m-1})=\M^{-m,m-1}\bigotimes
{\bf 1'}+ (k^{m-1}(k^m)^{-1})^{\f 12}\bigotimes \M^{-m,m-1}
\cr}~~~~~~~~~~~~~~~~
m\ge h(N)+1
\ee
(${\bf 1'}$ here denotes the unit of $Dif\!f({\bf R}_q^N)$, which acts
as the identity when considered as an operator on $Fun({\bf R}_q^N$),
and $\k^0\equiv {\bf 1'})$,
and is extended to all of $U_q^N$ as an homomorphism.
The counit $\epsilon$ is an homomorphism defined by (see formulae (9),(11)):
\be
\epsilon(u)= u \1_{\FR}|~~~~~~~~u\in U_q^N,~~~u\equiv differential~operator
{}~on~\FR
\ee
which on the generators takes the form
\be
\epsilon(\M^{ij})=\M^{ij}\1_{\FR}|=0~~~~~~~~~~~~~~~~~~~~\epsilon(\k^i)=
\k^i\1_{\FR}|=1.
\ee
$\epsilon, \phi$ are matched so as to form a bialgebra. For instance, relation
(58) is a consequence of the Leibnitz rule
\be
\k^i(f\cdot g)|=(\k^i f|)\cdot (\k^i g|),~~~~~~~~~f,g\in \FR,
\ee
which is a consequence of commutation relations (54) and definition (60).
 The validity of the
axioms of a bialgebra follow from the properties of $\DFR$, in particular
the coassociativity of $\phi$ follows from the associativity of the Leibnitz
rule, which in turn is a consequence of the associativity of
$\FR$, and the fact that $\phi,\epsilon$ are algebra homorphisms follows
from the associativity of the whole $\DFR$.
Finally the antipode $\sigma$ is found by consistency with the coalgebra;
it is an antihomomorphism which on the generators takes
the form
\be
\sigma(\k^i)=(\k^i)^{-1},
\ee
\be
\cases{\sigma(\M^{1-m,m})=-(\k^m(\k^{m-1})^{-1})
^{\f 12}\M^{1-m,m}\cr
\sigma(\M^{-m,m-1})=-(\k^m(\k^{m-1})^{-1})
^{\f 12}\M^{-m,m-1}\cr}
{}~~~~~~~~~~~m\ge\cases{1~~~~if~~N=2n+1\cr 2~~~~if~~N=2n\cr}
\ee
The Hopf structure of the inhomogeneous extension $U_q(e^N)$ of $U_q^N$
will be determined by  exactly the same procedure.

\section{Euclidean Hopf algebra $U_q(e^N)$ of u.e.a. type}

\subsection{Construction of $U_q(e^N)$}

{}~~~In this section we are going to introduce a
q-deformed Euclidean Hopf algebra $U_q(e^N)$,
of u.e.a. type  in $N$ dimensions as an extension
of $U_q^N=U_{q^{-1}}(so(N))$, more precisely by adding to the generators
of the latter
`` infinitesimal '' generators $p^i$ of translations and a generator $\La$
of dilatations. $U_q(e^N)$  will be dual of both
the Hopf algebras $\FE$, $\FBE$ (see Chapter III) through two different
parings.

We first introduce the algebraic structure of the Hopf algebra
$U_q(e^N)$.

The dual of a copy $\rn_{q,\vec{x}}^N$
of the Euclidean quantum space (thought as the braid group of finite
translations) is another copy $\rn_{q,\vec{p}}^N$ of this
quantum space, with generators $p^i$ such that $<p_i,x^j>=\delta^i_j$,
and $\FS$ coacts on it in the same way (see formula (33)),
provided we use the controvariant components $p^i=C^{ij}p_j$.

The condition that the $p^i$ generate $Fun(\rn^N_{q,\vec{p}})$
of course means that they satisfy the commutation relations
\be
{\cal P}_{A~hk}^{~~ij}p^h p^k =0.
\ee
The dual version of the statement that $\FS$ coacts on
$Fun(\rn^N_{q,\vec{p}})$ in the same way as on
$Fun(\rn^N_{q,\vec{x}})$ is that the commutation relations
between the generators of $U_q^N$ and $p^i$ must be the same
as those with $x^i$ (this is no surprise since we already saw that
also the derivatives $\p^i,\bp^i$ satisfy commutation relations (54)-
(56)).
We rewrite them here for convenience.
Let $m\ge h(N)+1$. Then:
\be
\cases{
[\M^{1-m,m},p^{m-1}]_q=-q^{\rho_m}p^m~~~~~~~~~~~~~~~
[\M^{1-m,m},p^{-m}]_q=q^{\rho_m}p^{1-m}, \cr
[\M^{-m,m-1},p^m]_{q^{-1}}=-q^{\rho_m}p^{m-1}~~~~~~~~~~
[\M^{-m,m-1},p^{1-m}]_{q^{-1}}=q^{\rho_m}p^{-m}, \cr
[\M^{01},p^0]=-q^{-1}p^1~~~~~~~~~~[\M^{-1,0},p^0]=p^{-1}~~~~~~~~~~~~~~~~~
if~~~N=2n+1
\cr}
\ee
\be
\cases{
[\k^i,p^h]_{a_{i,h}}=0,~~~~~~~~\cr
[\M^{1-m,m},p^i]_{b_{m,i}}=0~~~~~~~~
[\M^{-m,m-1},p^i]_{b_{m,i}}=0~~~~~~~~for~the~remaining~pairs~(m,i) \cr
\cr},
\ee

We see that the algebra generated by $\M,\k,p$ is closed.

To find a coproduct we will need to introduce
one more generator, the generator $\La$ of dilatation, such that:
\be
[\La,\vec{p}]_{q^{-1}}=0~~~~~~~~~~~~[\La, \k]=0~~~~~~[\La,\M]=0;
\ee
we will see that it is the dual of $w$ of section II.2.

Note that all the commutation relations involving $p$, (65)-(68),
are homogeneous in these generators.

{\bf Definition} In the sequel $\hat u_q(e^N)$ will denote the
algebra generated by $\M,\k,p$, $u_q(e^N)$ the algebra generated
by $\M,\k,p,\La^{\pm 1}$; the generators $\M,\k$ satisfy relations
(46)-(52).

$U_q^N\subset \hat u_q(e^N)\subset u_q(e^N)$ as subalgebras; more
precisely these subalgebras  can be projected into each other by
setting $\La=1$ and $p=0$ respectively in formulae (65)-(68).

{\bf Remark} Note that there exists a natural embedding
$\hat u_q(e^N)\hookrightarrow\hat u_q(e^{N+2})$ obtained
by setting equal to zero all the generators of $p^i,\M^{ij},\k^i$ of
$\hat u_q(e^{N+2})$ where either $i$ or $j$ takes the values $\pm (n+1)$.

Actually as a set of algebraically independent generators of $u_q(e^N)$
one should take the Chevalley generators of $U_q^N$, $\La^{\pm 1}$
and $one$ particular
momentum component (e.g. $p_n$), since equations (66) can be used to
$define$ the others; then relations (67) will be linear in $p$ and of
degree $d\ge 1$ in the Chevalley generators and will play a role
analogous to Serre relations (50)-(52).
For instance, with the above choice $p_n=\p_n$ the new `` Serre '' relations to
be added to the $U_q^N$ ones would read
\be
[\La,\p_n]_{q^{-1}}=0~~~~~~~~~~~~[\La, \k]=0~~~~~~[\La,\M]=0;
\ee
$$
\k^i\p_n=q^{-2\delta^i_n}\p_n\k^i~~~~~~~~~~~~
[\M^{-m,m-1},\p_n]=0=[\M^{1-m,m},\p_n]~~~~~~~~~ h(N)+1\le m<n
$$
\be
[\M^{-n,n-1},\p_n]_q=0~~~~~~~~~~~~~~\left[\M^{1-n,n},[\M^{1-n,n},\p_n]_q
\right]_{q^{-1}}=0.
\ee

{}~~

The commutation relations among $\M,\k,p,\La$ have been determined
by realizing these generators as differential operators on $\FR$;
now we can think of them as abstract generators of an algebra,
i.e. no more as elements of $Dif\!f(\rn_q^N)$.
At the representation-theoretic level this will be necessary since
we are interested in finding all
representations of $u_q(e^N)$ and not only those with zero $U_q(so(N))$-spin.

We try now to endow $u_q(e^N)$ with some Hopf structures compatible
with its algebraic structure.

We first look for the coalgebra structure. Again, as we
did in section II.4  with $U_q^N$, our postulates
will be suggested by the explicit realization of the generators $p^i$ as
differential operators on $Fun(\rn_q^N$).

A first possibility is to realize $p^i$ by $p^i\equiv \p^i$, a second,
by $p^i\equiv \bp^i$. Using relations (18),(22),(39)
one can easily show that
more generally $p^i\equiv f(\La)(\alpha\p^i+\beta\bp^i)$ is a realization of
the $p^i$'s as differential operators on $\rn_q^N$. The following theorem
shows that there are no more alternatives.

\begin{theorem}
The only way to realize the generators $p^i$ satisfying the algebra
(65)-(68) as
elements of $Dif\!f(\rn_q^N)$ is by a combination
\be
p^i\equiv f(\La)(\alpha\p^i+\beta\bp^i),~~~~~~~~~~~~\alpha,\beta\in\cn,~~~
f(t)\in \cn[t]
\ee
\end{theorem}

$Proof$. The commutation relations (66)-(67) impose for $p^i$ the general
form $p^i=S_1\p^i+S_2 x^i$, where $S_1,S_2\in Dif\!f(\rn_q^N)$
are scalars. The commutation relation $(68)_1$ implies
that the natural dimensions $d$ of $S_1,S_2$ are given by $d(S_1)=0$,
$d(S_2)=2$. Hence, using the results mentioned ate the end of section II.1,
we are led to an improved ansatz
$p^i=S_1'(B,\La)\p^i+S_2'(B,\La)x^i\p\cdot\p$ (recall that $B\in U_q^N$).
Now we replace this general form for $p^i$ into the commutation relations
(65).
We move $\p \cdot \p$ to the right of $S_1',S_2',x^i,\p^i$ by using relations
(18),(21),(23).
The three coefficients of the powers 0,1,2 of $\p \cdot \p$ must vanish
independently:
\be
\cases{{\cal P}_{A~hk}^{~~ij}S'_1(B,\La)\p^hS'_1(B,\La)\p^k=0 \cr
{\cal P}_{A~hk}^{~~ij}[S_1'(B,\La)\p^hS_2'(B,\La)x^k+S_2'(B,\La)x^h
S_1(B,\La q^2)\p^k+q^{2-N}S_2x^hS_2(B,\La q^2)\p^k]=0 \cr
{\cal P}_{A~hk}^{~~ij}S_2'(B,\La)x^kS_2'(B,\La q^2)x^k=0. \cr}
\ee
We expand $S_i'$ in a power series
$S_i'=\sum\limits_{n=0}^{\infty}B^nS_{i,n}'(\La)$ and use the commutation
relations (40) to move the $\p$ (resp. the $x$) to the right of all the $B$'s
in the first (resp. third) equation; we get an expansion in powers of the
independent generators $B,l^{ij}$ of $U_q^N$. Setting all their coefficients
equal to zero implies $S_{i,n}'\equiv 0$ $n\ge 1$
(only the coefficient of the constant term vanishes automatically because
of the relation (7). In other words $S_i'=S_i'(\La)$ only.
This allows to rewrite the second equation as
\be
[-q^2S_1'S_2'(\La q)+S_2'S_1'(\La q)+S_2'S_2'(\La q)q^{2-N}]l^{ij}\p\cdot\p=0;
\ee
This implies that the term in square brackets must vanish.
A nontrivial solution of the equations (72) is therefore $S_2'=0$,
$S_1'=S_1'(\La)$, yielding the solution with $\beta=0$ given in the claim.
If $S_2'\neq 0$ we can set $s(\La):=\f{S_1'(\La)}{S_2'(\La)}$, and
the preceding equation is equivalent to the new one
\be
q^{2-N}-q^2s(\La)+s(q\La)=0.
\ee
Expanding $s(\La)$ in power series one immeditely finds that the
general solution of the latter equation is
$s=\f{q^{2-N}}{q^2-1}+a\La^2$, $a\in\cn$, which yields the remaining
solutions of the claim (after use of eq. (72)). $\diamondsuit$

{\bf Remark} Note that $p^i\in u_q(e^N)$
for all realizations (71) of $p^i$; for, if e.g. we had picked
$\p^i$ as original realizations of $p^i$, it is easy to check using
formulae (19),(22),(23),(40)
that
\be
\bp^i =\f{1+q^{N-2}}{1-q^{-2}}\La^{-1}[B,\p^i]_{q^{-1}}=
\La^{-1}(B\delta^i_b+\f{q^4-1}{q^{4-N}+1}l^{ic}C_{cb})\p^b.
\ee

Now we can use theorem 2 to
derive the natural coalgebra structure associated to the
realization $u_q(e^N)\subset Dif\!f(\rn^N_q)$, and then convert it into
a Hopf one by finding the antipode.
We follow the same approach used in section II.4 with $U_q^N$, i.e. the
counit derives from evaluation of differential operators on the
unit ${\bf 1}\in\FR$, the coproduct from their Leibnitz rules, and the
antipode from consistency with the coproduct, according
to the axioms of a Hopf algebra.

Because of the remark following theorem 1, it is sufficient to show
explicitly the exixtence and the form of
coproduct and counit $\phi,\epsilon$ only on the generators
$p^i=\p^i$; even better,
only on one particular `` momentum '' component, say $\p_n$. In fact, we
can determine the action of $\phi,\epsilon$ on the other components $\p^i$
iteratively through commutators (66) of $\p_n$
 with the Chevalley generators of $U_q^N$, and on the
other realizations (71) by using relation (75). In both steps, we use
the basic property that $\phi,\epsilon$
are algebra homomorphisms.

 From
\be
[\La,\vec{x}]_q=0
\ee
and
\be
\p_nx^n=1+q^2x^n\p_n,~~~~~~~~~~~~~~~~~~\p_nx^{-n}=x^{-n}\p_n,
{}~~~~~~~~~~~~~~~~~~\p_nx^l=qx^l\p_n,~~~~~|l|<n
\ee
we infer
\be
\phi(\La):=\La\ot\La
{}~~~~~~~~~~~~\epsilon(\La)=1~~~~~~~~~~~~~~~~
\sigma(\La)=\La^{-1}
\ee
and
\be
\phi(\p_n):=\p_n\ot\1'+\La(\k^n)^{\f 12}\ot\p_n
{}~~~~~~~~~~~~\epsilon(\p_n)=0~~~~~~~~~~~~~~~~
\sigma(\p_n)=-\La^{-1}(\k^n)^{-\f 12}\p_n.
\ee
$\phi, \epsilon$ are extended as algebra homomorphisms, $\sigma$ as an algebra
antihomomorphism, to the rest of $u_q(e^N)$.
Then $u_q(e^N)$ is turned into a Hopf algebra, what we call $U_q(e^N)$,
since the basic axioms are satisfied on the generators.
In particular we readily find when $i\ge h$, for instance,
\be
\phi(\p_i)=\p_i\ot {\bf 1'}+ \La(\k^i)^{\f 12}\ot \p_i
+(1-q^2)\sum\limits_{j>i}q^{-\rho_i}\La\M^{-i,j}(\k^j)^{\f 12}\ot \p_j
{}~~~~~~~~~~~~\epsilon(\p_i)=0
\ee
\be
\sigma(\p_i)=-\La^{-1}(\k^i)^{-\f 12}\p_i+(q^2-1)
\sum\limits_{j>i}q^{-\rho_i}\La^{-1}(\k^j)^{-\f 12}\sigma(\M^{-i,j})\p_j.
\ee

As for the choice of `` conjugated '' translation generators
$p^i\equiv\bp^i$, in principle the actions of $\phi,\epsilon,\sigma$
can be determined from formula (75) in the remark, but
practically it is more convenient to find them using the same procedure
as for $p^i=\p^i$. Only, we pick now as initial
component $p_{-n}\equiv\bp_{-n}$.
{}From
\be
\bp_{-n}x^{-n}=1+q^{-2}x^{-n}\bp_{-n},~~~~~~~~~~~~~~~~~~\bp_{-n}x^n=x^n\bp_{-n},
{}~~~~~~~~~~~~~~~~~~q\bp_{-n}x^l=x^l\bp_{-n},~~~~~|l|<n
\ee
we find
\be
\phi(\bp_{-n}):=\bp_{-n}\ot\1+\La^{-1}(\k^n)^{\f 12}\ot\bp_{-n}
{}~~~~~~~~~~~~\epsilon(\bp_n)=0=\epsilon(\bp_i)~~~~~~~~~~~~~~~~
\sigma(\bp_{-n})=-\La(\k^n)^{-\f 12}\bp_{-n}.
\ee

So far we have not considered the $*$-structure. Of course,
we would be finally interested in a $*$-Hopf algebra,
i.e. in a Hopf algebra endowed with a $*$ operation which is compatible
with the Hopf axioms.
Whenever $q\in \rn^+$, the $*$-structure of $\DFR$ induces
at the algebra level a consistent $*$-structure for $U_q(e^N)$, see formulae
(13),(14),(53): in other words, $u_q(e^N)$ by itself is a $*$-algebra.
We can select a particular realization (71) of the momenta $p^i$ by
imposing the same reality condition valid for the $x^i$,
\be
(p^i)^*=p^jC_{ji}.
\ee

\begin{corollary}
The only way to realize the generators $p^i$ satisfying the algebra
(65)-(68) and the
$*$-relations (84) as
elements of $Dif\!f(\rn_q^N)$ is (up to a global real factor) through
\be
p^i\equiv -i(\p^iq^N+\bp^i).
\ee
\end{corollary}

However, this is not of much help, since

\begin{prop}
The above $*$-structure is incompatible with the Hopf one.
\end{prop}
$Proof$. That $*$ and $\phi$ do not commute can be immediately
checked on $\La$:
\be
[(*\ot *)\circ\phi](\La)=q^{-2N}\La^{-1}\ot \La^{-1}\neq
q^{-N}\La^{-1}\ot \La^{-1}=\phi(\La^*). ~~~~~~\diamondsuit
\ee
In other words, $U_q(e^N)$ is not a $*$-Hopf algebra.
This is a major problem for physical applications, since it prevents
the standard construction of Hilbert space representations of
composite physical systems through tensor product of Hilbert space
representations of fundamental ones, and is common to many
inhomogeneous q-deformed Hopf algebras.

We think that the way out of this problem should be seeked in
a non-standard way to perform tensor products of Hilbert spaces.

The $*$-algebra structure
is all what we will use to find the fundamental (i.e. one-particle)
Hilbert space representations of $u_q(e^N)$ (see Chapter IV).
On $p^i$ we will then impose the $*$-relations (84) at an abstract level.
However, to find a realization of the singlet (i.e. spin-zero)
representation on
`` $\rn^N_q$ configuration space '', i.e. realize the carrier space of
this representation as a subspace of $\FR$, we will use both $\p^i$ and $\bp^i$
in a non-standard way (section (IV.2).

Summing up,
our definition of the quantum Euclidean Hopf algebra (of u.e.a. type)
$U_q(e^N)$ can be taken as follows

{\bf Definition:} $U_q(e^N)$ as an algebra is
generated by elements $\La^{\pm 1}\k^i,\M^{-i,i+1},\M^{-i-1,i}$
($i=h,h+1,...,n$) and
$p_n$ (in the $N=2n$ case one should also add $\M^{1,2},\M^{-2,-1}$).
They satisfy the commutation relations (46)-(52), (65)-(68).
The Hopf structure is given by (58)-(61),(63),(64),(78),(79).

\subsection{The dual Hopf algebras of $U_q(e^N)$}

In this section we recognize that $U_q(e^N)$ is the dual Hopf algebra of both
versions
$Fun(E^N_{q^{-1}})$, $Fun(\bar E^N_{q^{-1}})$ of the Euclidean Hopf algebra
of functions-on-the-group type introduced in Ref. \cite{schl},
and reviewed in Chapter III. This means, incidentally, that suitable
completions of $\FE$, $\FBE$ will coincide.

To prove that statement one has to find a pairing $<~~,~~>$ between
$U_q(e^N)$ and $Fun(E_{q^{-1}}^N)$ (resp. $Fun(\bar E_{q^{-1}}^N)$)
satisfying linearity and the two postulates
\be
<ab,A>=<a\ot b,\phi^E(A)>~~~~~~~~~
<a,AB>=<\phi(a),A\ot B>
\ee
$a,b\in U_q(e^N),~~~A,B\in \FE~~~(resp.~~\in~~\FBE),$
where
\be
<a\ot b,A \ot B>:= <a,A><b,B>.
\ee
Since $\FE,\FBE$ are not dual quasitriangular Hopf
algebras (implying that the basic commutation relations between
all its generators cannot be put in the form $(26)_1$, with a suitable
$\hat R$ matrix), the pairing cannot be introduced through a big $\hat R$
matrix as in the homogeneous case. Nevertheless, it is natural
(and, as we are going to see, correct) to introduce it as an extension of
the pairing between $U_{q^{-1}}(so(N))$ and $Fun(SO_{q^{-1}}(N))$ exhibited
in Ref. \cite{frt}.

It is sufficient to exhibit such a pairing on a pair of sets of generators
of the two Hopf algebras, since then it can be extended in a
consistent way from the generators to the
rests of the algebras.

As a set of generators of $U_q(e^N)$ we choose the Chevalley generators $\M,\k$
of $U_q^N$ together with $\p_n,\La^{\pm 1}$ (resp. $\bp_{-n},\La^{\pm 1}$).
As a set of generators of
$Fun(E_{q^{-1}}^N)$ (resp. $Fun(\bar E_{q^{-1}}^N)$) we choose the
generators $w^{\pm 1},T^i_j$ together with $y^i$ (resp. $\bar y^i$)
(see section III.4). Thus, our problem is reduced to exhibiting
a pairing such that
the $U_q^N$ algebra together with relations (69,(70) (resp.
the analogue for $\bp_{-n}$) is compatible with the coalgebra structure
of $Fun(E_{q^{-1}}^N)$ (resp. $Fun(\bar E_{q^{-1}}^N)$), and the algebra
relations (26)-(28) of $Fun(E_{q^{-1}}^N)$ (resp. $Fun(\bar E_{q^{-1}}^N)$)
are compatible with the coalgebra of $U_q(e^N)$.

\begin{prop}
$U_q(e^N)$ id dual of $Fun(E_{q^{-1}}^N)$, $Fun(\bar E_{q^{-1}}^N)$
respectively through the natural pairings
\be
\cases{
<\La,w^{\pm 1}>=q^{\pm 1},~~~~~~~~<\La,T^i_j>=\delta^i_j,~~~~~~~~<\La,y^i>=0
\cr
<\p_n,w^{\pm 1}>=0~~~~~~~~<\p_n,y^j>=\delta_n^j,~~~~~~~<\p_n,T^i_j>=0 \cr
{}~~~~~~~~<\M^{ij},w>=0~~~~~~~~<\M^{ij},y^k>=0
{}~~~~~~~~<\k^i,w^{\pm 1}>=1~~~~~~~~~~<\k^i,y^j>=0 \cr
<U_q^N,Fun(SO_{q^{-1}}(N))>: the~pairing~of~Ref.~\cite{frt}~with~q\rightarrow
q^{-1} \cr}
\ee
and
\be
\cases{
<\La,w^{\pm 1}>=q^{\mp 1},~~~~~~~~<\La,T^i_j>=\delta^i_j,~~~~~~~~<\La,\bar
y^i>=0
 \cr
<\bp_{-n},w^{\pm 1}>=0~~~~~~~~<\bp_{-n},\bar y^j>=\delta_{-n}^j,~~~~~~~
<\p_{-n},T^i_j>=0 \cr
{}~~~~~~~~<\M^{ij},w^{\pm 1}>=0~~~~~~~~<\M^{ij},\bar y^k>=0
{}~~~~~~~~<\k^i,w>=1~~~~~~~~~~<\k^i,\bar y^j>=0 \cr
<U_q^N,Fun(SO_{q^{-1}}(N))>: the~pairing~of~Ref.~\cite{frt}~with~q\rightarrow
q^{-1}. \cr}
\ee
\end{prop}

Of course the pairing of the generators with the units is given by
$<{\bf 1'},A>:=\ve(A)$,

$<a,{\bf 1}>:=\epsilon(a)$,
{}~$a,{\bf 1'}\in U_q(e^N)$, $A,{\bf 1}\in Fun(E_{q^{-1}})$ and $\ve$ is the
counit in $Fun(E_{q^{-1}}^N)$.

$Proof$: See the Appendix $\diamondsuit$

As a consequence of the above definition, it is easy to show that the
pairing between $U_q^N$ and the subalgebra $P$
of translation generators extends in a trivial way, $<P,U_q^N>=0$,
whereas the pairing between the latter and the generators $y,\bar y$ of
finite translations extends first as
\be
<\p_i,y^j>=\delta_i^j=<\bp_i,\bar y^j>,
\ee
and then as
\be
<\p_{i_1}\p_{i_2}...\p_{i_l},y^{j_1}y^{j_2}...y^{j_m}>=
\cases{0~~~~if~~m\neq l \cr
([1;\hat R][2;\hat R]...[m;\hat R])
^{j_1j_2....j_m}_{i_mi_{m-1}...i_1}
{}~~~~if~~m=l, \cr}
\ee
for the unbarred objects and for the barred ones by replacing
$\hat R \rightarrow \hat R^{-1}$. Here, (with a notation suggested by, but
slightly different from that of Ref. \cite{majf})
\be
[h,M]:=1+M_{h-1,h}+M_{h,h+1}M_{h-1,h}...+
M_{m-1,m}M_{m-2,m-1}...M_{h-1,h}.
\ee
Note that formulae (92) alone can be obtained also by using the fact that
the braid group $\rn_{q,\vec{\p}}^N$ (resp.
$\rn_{q,\vec{\bp}}^N$) of momenta and the braid group
$\rn_{q,\vec{x}}^N$ of finite translations $x$ are
`` braided '' paired, in the sense that their pairing is generated from the
basic formulae (91),(87), provided $\phi^E$ is replaced by the braided
coaddition $\und{\Delta}$ (in the language of Ref. \cite{maj2}),
$\und{\Delta}(x^i):=x^i\uot {\bf 1}+{\bf 1}\uot x^i$ and $\phi$ is replaced by
a braided one $\und{\Delta}'$, $\und{\Delta}'(p^i):=p^i\uot {\bf 1}+
{\bf 1}\uot p^i$.

{}From the viewpoint of duality the difficulty one meets
in finding $*$-structures compatible with
the Hopf ones in
$\FE,U_q(e^N)$,$\rn_{q,\vec{\p}}^N$,$\rn_{q,\vec{x}}^N$ (resp.
$\FBE,U_q(\bar e^N)$,$\rn_{q,\vec{\bp}}^N$,$\rn_{q,\vec{\bar x}}^N$)
are all related
and are a
general feature of inhomogeneous quantum groups of BWM
type \cite{bwm}.

\subsection{New $L$ generators of the Euclidean algebra $u_q(e^N)$}

{}~~~In Chapter IV we will construct the fundamental Hilbert space
representations of the Euclidean algebra. We will show (Proposition 7)
that for such representations either $(p\cdot p)_i\equiv 0$
identically, $\forall i\ge h$, or all $(p\cdot p)_i$ are strictly
positive definite.
In the former case the algebra reduces to the homogeneous one $U_q^N$,
in the latter case, which we here consider, it follows that we can define
the inverse of $(p\cdot p)_i$.

Let us consider the relations
\be
[\M^{-m,m+1},(p\cdot p)_l]=0~~~~~~~~~~~~~~~~~~
if~~~l\neq m;
\ee
\be
[\M^{-m,m+1},(p\cdot p)_m]=-q^{2\rho_{m+1}}p^{-m}p^{m+1},~~~~~~~~~~
(\Rightarrow~~~~~[\M^{-m,m+1},\f{1}{(p\cdot p)_m}]=
\f{q^{2+2\rho_{m+1}}}{[(p\cdot p)_m]^2}p^{-m}p^{m+1}),
\ee
which can be easily drawn from equations (66),(67).
As a consequence of the second
formula, $\M$ would not map eigenvectors of $(p\cdot p)_m$ into each-other.
However, one can define improved generators $L$ which actually do.
{}From (67) we find
\be
\cases{
[\M^{-m,m+1},\f{1}{(p\cdot p)_m}]=\f{q^{2\rho_{m+1}+2}}
{[(p\cdot p)_m]^2}p^{-m}p^{m+1},\cr
[\M^{1,2},\f{1}{(p\cdot p)_1}]=\f{1}
{[(p\cdot p)_1]^2}p^1p^2,~~~~~~~~~~~~if~~~~N=2n;\cr}
\ee
therefore we define
\be
\cases{
L^{-m,m+1}:=\M^{-m,m+1}+\f{q^{2\rho_{m+1}+2}}{(1-q^2)(p\cdot p)_m}p^{-m}
p^{m+1}\cr
L^{-m-1,m}:=\M^{-m-1,m}+\f{q^{2\rho_{m+1}+1}}
{(1-q^2)(p\cdot p)_m}p^{-m-1}p^m. \cr}
\ee
(similarly for $L^{12}$). Note that this redefinition is possible only
when $q\neq 1$. The basic property of the new generators is the fact that
if $i>0$ then (see definition (57))
\be
[L^{-m,m+1},p^i]_{b_{i,m}}=0~~~~~~~~~~~~[L^{-m-1,m},p^i]_{b_{i,m}}=0,
\ee
implying
\be
[L^{-m,m+1},(p\cdot p)_i]=0=[L^{-m-1,m},(p\cdot p)_i]~~~~~~~~~~\forall i,m;
\ee
moreover, it is easy to see
that the $L$'s satisfy the same *-conjugation relations as the $\M$'s.

Let us find out now the commutation relations satisfied by the $L$'s.
We can define other roots $L$ starting from
simple ones, just in the same way as we did with the $\M$'s,
using relations (43)-(45) (with the replacement $\M\rightarrow L$).
$L$ roots can be divided into positive and negative ones according to the
same convention used for the $\M$'s.

\begin{prop} Let $k\ge h+1$. The commutation relations between positive and
negative simple roots are
\be
[L^{1-m,m},L^{-k,k-1}]_a=0~~~~~~~~~~~~~~~a=\cases{q^{-1}~~~~m\pm 1=k \cr
1~~~~if~~k\neq m, m\pm 1, \cr}
\ee
\be
[L^{1,2},L^{-2,1}]=\f{(p\cdot p)_2p^1p^1}{(1-q^2)[(p\cdot p)_1]^2},
{}~~~~~~~~
[L^{-1,2},L^{-2,-1}]=\f{(p\cdot p)_2p^{-1}p^{-1}}{(1-q^2)[(p\cdot p)_1]^2},~~
{}~~~~~~~~~~~~~if~~~N=2n,
\ee
\be
\cases{[L^{1-m,m},L^{-m,m-1}]_{q^2}=q^{1+2\rho_m}\f{1-\k^{m-1}(\k^m)^{-1}}
{q-q^{-1}}+C_m~~~~~~~~2\le m\le n\cr
[L^{01},L^{-1,0}]_q=q^{-\f 12}\f{1-(\k^1)^{-1}}{q-q^{-1}}+C_1
{}~~~~~~~if~~N=2n+1\cr}
\ee
where
\be
C_1:=\f{q^{\f 12}}{1-q^2}\left[1+q\f{(p\cdot p)_1}{(p\cdot
p)_0}\right]~~~~~~~~~
if~~~N=2n+1
\ee
\be
C_{m+1:}=\f{q^{2\rho_m}}{1-q^2}\left[1-\f{(p\cdot p)_{m-1}(p\cdot p)_{m+1}}
{[(p\cdot p)_m]^2}\right],~~~~~~~~~~~m\ge 1.
\ee
\end{prop}
$Proof$. Use equations (47)-(48),(96),(99)
perform explicit computations.
$\diamondsuit$

The list of commutation relations involving the $L$'s is completed
by the following proposition:
\begin{prop}
The $[\k,L]$ relations, the Serre relations and the $*$-relations
for the $L$ generators are
the same as those of the $\M$ generators
\end{prop}
$Proof$: explicit computations. $\diamondsuit$

Summing up, the commutations relations among the $L$'s are the same as those
among the $\M$'s, if we add some `` central charges '' ($C_m$).

{\bf Remark} Note that the embedding
$\hat u_q(e^N)\hookrightarrow\hat u_q(e^{N+2})$ mentioned after formula (68)
is obtained  now by setting equal to zero all the generators
$p^i,L^{ij},\k^i$ of
$\hat u_q(e^{N+2})$ having  $i=\pm (n+1)$ for some space index $i$.

{\bf Definition} (Borel decomposition)
We denote by $u_q^{\pm,N}$ the subalgebra of $u_q(e^N)$ generated by
the positive roots $L$'s (resp. by the negative roots $L$'s and, in the case
$N=2n$ only, by $p^{\pm 1}$).

\subsection{Casimirs of $\hat u_q(e^N)$}

As in the classical case, $\hat u_q(e^N)$ has $n+1-h$ casimirs;
their definition mimics the classical one.
They can be built in the most compact way as follows. Define the
`` Pauli-Lubanski '' $(2l+1)$ - form
\be
\omega_{2l+1}:=(\omega)^lp~~~~~~~~~~~\omega:=l^{ij}\xi_j\xi_i,~~~~~p:=p^i\xi_i
{}~~~~~~~~~~~~l=0,1,...,n-1.
\ee

\begin{theorem}
The $n+1-h$ casimirs $\Omega^l$ of $\hat u_q(e^N)$ are defined by
\be
\Omega^l~dV:= \omega_{2l+1}\wedge^{\s} \omega_{2l+1},
\ee
and
\be
\Omega^n~dV:=\omega_{2n+1}~~~~~~~~~~~~~~~~~~~if~~~~N=2n+1,
\ee
where $dV$ denotes the volume $N-form$, and $\s$ denotes the Hodge
duality operation introduced in section II.3
\end{theorem}
$Proof$. The general proof will be given in a separate paper \cite{fio7};
it involves the use of the general $[l^{ij},p^h]$ commutation relations
and of specific properties of the $q$-deformed completely antisymmetric
projector with $m\le N$ indices. The theorem will be verified
in the cases $N=3,4$ using the $L$ generators in next subsection.
$\diamondsuit$

The irreps of $\hat u_q(e^N)$  are characterized by the values of the casimirs.

In particular, when $l=0$ we find the square momentum casimir
\be
\Omega^0 \equiv (p\cdot p)_n;
\ee
when $N=4$, $\Omega^1$ is the q-deformed analogue of the Wick-rotated
casimir which is constructed from the Pauli-Lubanski vector $w^i$
defined by $w^i\xi_i:=^*\omega_3$ and
gives the intrinsic spin of each irrep. Using the properties
of the $\epsilon_q$ tensor, one can show that $w^i$ really generate a
quantum Euclidean space, i.e. satisfy
\be
{\cal P}^{~~ij}_{A_hk}w^hw^k=0.
\ee
Of course, the casimirs
$\Omega^l$, $l\ge 1$, are zero in the singlet representation, i.e. that
characterized by the trivial weight, since then both $p^l$ and
$l^{ij}$ can be expressed as differential operators on $\rn_q^N$,
namely $l^{ij}={\cal P}_{A~hk}^{~~ij}x^h\p^k\Lambda^{-1}$, $p^l=\p^l\Lambda^a$,
and ${\cal P}_{A~hk}^{ij}\p^h\p^k=0$.

\subsubsection{The casimirs of $\hat u_q(e^N)$ in the cases $N=3,4$ in terms
of the $L,\k,p$ generators}

\begin{prop}
When $N=3,4$, the Casimirs $\Omega_1$ (107), (106)
in terms of $p,L,\k$ generators take respectively the form
\be
\Omega_1=p^0(\k^1)^{-\f 12}-q(q+1)\f{(p\cdot p)_1}{p^0}(\k^1)^{\f 12}
+q^{\f 12}(1-q)(1-q^2)L^{-1,0}L^{0,1}(\k^1)^{\f 12}
\ee
and
$$
\Omega_1=(L^{-2,1}L^{-1,2})(L^{-2,-1}L^{1,2})\k^2(p\cdot p)_1
+\f{q^{-2}}{(q^2-1)^2}
(p\cdot p)_1\{\k^1(L^{-2-1}L^{12})+(\k^1)^{-1}(L^{-21}L^{-12})\}
$$
\be
+\f{q^{-4}(p\cdot p)_1(\k^2)^{-1}}{(q^2-1)^4}\left[1-q^2\k^2
\f{(p\cdot p)_2}{(p\cdot p)_1}\right]^2
-\f{q^{-2}(p\cdot p)_2}{(1-q^2)^2(p\cdot p)_1}[p^{-1}p^{-1}L^{-21}L^{12}
+p^1p^1L^{-2-1}L^{-12}]\k^2.
\ee
\end{prop}
$Proof$. We prove that $\Omega_1$ as defined by equation (107), (106)
take the forms (110),(111); then it is straightforward
to verify that they are casimirs
of $u_q(e^N)$ using the commuation relations of the preceding subsection.

Case $N=3$. Using the definition of ${\cal P}_A$ one verifies that there
are only three independent $l^{ij}$, $l^{-1,0},l^{01},l^{1,-1}$, say.
$l^{0,1}=(\k^1)^{\f 12}\M^{0,1}$,
$l^{-1,0}=(\k^1)^{\f 12}\M^{-1,0}$, $l^{1,-1}=\f{(\k^1)^{\f 12}-B}{q-1}$
see ref. \cite{fio4}. $B$ can be
easily expressed as a linear funtion of
the quadratic casimir ${\cal C}_{U_q^3}$
of $U_q^3$, which in terms of $\M,\k$ generators reads
\be
{\cal C}_{U_q^3}=q^{\f 32} \M^{-1,0}\M^{01}(\k^1)^{\f 12}+\f{[(\k^1q)^{\f 14}-
(\k^1q)^{-\f 14}]^2}{(q-q^{-1})(q^{\f 12}-q^{-\f 12})};
\ee
one finds
\be
B(1+q)=(\k^1)^{\f 12}+\k^{-\f 12}+q^2(q-q^{-1})(q^{\f 12}-q^{-\f 12})
\M^{-1,0}\M^{01}(\k^1)^{\f 12}.
\ee
This allows an expression for $\Omega_1$ involving only $\M,\k$ generators.
Finally, one replaces $\M$'s as functions of $\k^1,L^{0,1},L^{-1,0}$ and
finds the expression (110).

The case $N=4$ can be proved in a similar way \cite{fio7}.  $\diamondsuit$

{}~~~~~~~~~~~~~~~~~~~~~~~~~~~~~~~~~~

\section{The fundamental Hilbert space representations of $u_q(e^N)$}

\subsection{Construction}

As noticed before, when $q\in\rn$, $u_q(e^N)$ is a $*$ algebra,
with $*$-relations (14),(25),(53).
We remind that a $*$-representation $\Gamma$ \cite{schmu}
of a $*$-algebra $A$ on a Hilbert ${\cal H}$ space
(briefly: Hilbert space representation) is essentially a
representation of $A$ such that $\Gamma(a^*)=\Gamma(a)^{\dagger}$
($T^{\dagger}$ is the adjoint of $T$) at least on a dense subset of the
Hilbert space. In this section we will determine the fundamental
representations of $u_q(e^N)$. We will find a basis of ${\cal H}$
and show how the generators of $u_q(e^N)$ are to be represented as operators
on the elements of the basis. We will not deal with questions regarding
domains of definition of the operators; this is premature at this
stage, and is out of the scope of this work.
The positivity of the scalar product
\be
\cases{~~~~(u,u)\ge 0,~~~~~, \cr (u,u)=0~~~\Leftrightarrow u=0, \cr}
{}~~~~~~~~~~~~~~~~~~~~\forall u\in{\cal H}
\ee
will be imposed $apriori$ at each step of our construction, and of course will
be essential in determining the structure of the representations.

\subsubsection{Spectra and eigenspaces of the squared momentum observables;
the action of $p,\k,L$ on the points of the `` q-lattice ''}

{}~~~Contrary to the classical case, the momenta $p^i$ don't commute with
each-other, therefore cannot be chosen
as (part of) a set of commuting observables in order
to study the Hilbert spaces of the irreps of $u_q(e^N)$. On the contrary,
among the commuting observables of a complete set characterizing an
irrep we can always take (see sections II.4, III.1)
\be
p_0,(p\cdot p)_1,...,(p\cdot p)_{n-1},(p\cdot p)_n;
\k^1,...,\k^n~~~~~~~~~~~~~~~~~~(p_0\equiv 0~~~if~~N=2n)
\ee
(in fact we will see that they actually make up a complete set for
the `` singlet '' irrep).
Notice that if we had taken $(p\cdot p)_0$ instead of $p_0$ we could not
distinguish between positive and negative eigenvalues of $p_0$.
It is easy to realize from the commutation relations of $u_q(e^N)$ that the
sign of $p_0$ will be the same within each irrep.

We make an ansatz, assuming existence of eigenspaces
of the first $n+h$ observables consisting only of normalizable eigenvectors;
then we find that ${\cal H}$ consists of eigenspaces of
normalizable eigenvectors, too.

Let ${\cal H}$ be the Hilbert base space of the considered irrep of $u_q(e^N)$.
Given an eigenspace $\hat {\cal H}\in {\cal H}$ of $(p\cdot p)_n$ with
eigenvalue $M^2$, $\hat{\cal H}_{\pi_n}:=\La^{2\pi_n+2}\hat {\cal H}$
($\pi_n \in \zn$) will be an eigenspace as well, with
eigenvalue $M^2q^{2\pi_n+2}$. Here $M^2$ is a
nonnegative constant characterizing the irrep;
it has the dimension of a mass squared and is determined up to integer
powers of $q^2$. Each $\hat{\cal H}_{\pi_n}$ will be an irrep of
$\hat u_q(e^N)$. Finally, the irrep of $u_q(e^N)$ characterized by $M=0$,
i.e. $(p\cdot p)_n\equiv 0$, is actually an irrep of $U_q^N$.

Therefore our problem is thus reduced to the study of the
irreps of $\hat u_q(e^N)$.

\begin{prop}
There are only the following two alternatives in ${\cal H}$:
\be\cases{
1)~~(p\cdot p)_i\equiv 0~~~~~~~identically~~~~\forall i=h,h+1,...,n; \cr
2)~~(p\cdot p)_i>0~~~~~~~strictly~~~~~\forall i=h,h+1,...,n . \cr}
\ee
\end{prop}
$Proof$. As a trivial observation, $(p\cdot p)_i\ge 0$, since
$p^jp_j= q^{2\rho_j}(p_j)^*p_j$. Then $(p\cdot p)_n\equiv 0$ implies
$(p\cdot p)_i\equiv 0$.

Suppose now that $(p\cdot p)_m>0$ (starting from $m=n$),
and assume ($per~absurdum$) that there
exists $|\phi>\in {\cal H}$ such that $(p\cdot p)_{m-1}|\phi>=0$. Formulae
(15),$(16)_2$ then imply $p^{\pm(m-1)}|\phi>=0$, $p^m|\phi>\neq 0$. Because of
relation (66), this in turn implies
\be
p^{m-1}\M^{1-m,m}|\phi>\propto [\M^{1-m,m},p^{m-1}]_q|\phi>
\propto p^m|\phi>\neq 0,
\ee
and, by taking the norm and using relations (66),(67),$(16)_1$,
the contradiction:
$$
0\neq<\phi|\M^{-m,m-1}p^{1-m}p^{m-1}\M^{1-m,m}|\phi>=
$$
\be
<\phi|\M^{-m,m-1}(\M^{1-m,m}p^{1-m}p^{m-1}+q^{\rho_m} p^mp^{1-m}|\phi>=0.
{}~~~~~~\diamondsuit
\ee

In the sequel $M^2>0$. From $(16)_2$ we find
\be
(p\cdot p)_l=p^lp^{-l}q^{\rho_l}+q^{-2}(p\cdot p)_{l-1}=
p^{-l}p^lq^{\rho_l}+(p\cdot p)_{l-1}
\ee
Each $p^lp^{-l}$ is positive definite
($\Rightarrow (p\cdot p)_l> (p\cdot p)_{l-1}> 0~~~\forall l$).

 Assume that $|\psi>\in {\cal H}$ is an eigenvector
of all $(p\cdot p)_i$. According to eq. (16),
$|\psi_{l,\pm r}>:=(p^{\pm l})^r|\psi>$
($r\in\nn,~~l\le n$) will also be an eigenvector of all
of them; the eigenvalues of $|\psi>,|\psi_{l,\pm r}>$ will differ by an
integer power of $q$. Let $a_i$ be the eigenvalues of $(p\cdot p)_i$
on $|\psi>$.
The norm of $|\psi_{l,-r-1}>$ will be given by
\be
<\psi_{l,-r-1}|\psi_{l,-r-1}>=<\psi_{l,-r}|\psi_{l,-r}>(a_l-q^{-2r-2}a_{l-1})
\ee
If $a_{l-1}\neq 0$,
there must exist a $r$ such that $(p^{-l})^{r+1}|\psi>=0$, otherwise the above
norm would get negative for large $r$. In other words there must
exist a state which is annihilated by $p^{-l}$.

Let us start by setting $l=n$ in the previous argument;
then we iterate it by setting $l=n-1,...,h+1$.
We infer the existence of a nonempty subspace
${\cal H}_{\pi_n,0}\subset \hat{\cal H}_{\pi_n}$ such that
$(p\cdot p)_n=M^2q^{2\pi_n+2}$,
$p^{-l}{\cal H}_{\pi_n,0}=0~~~\forall l$.
Let $\Pg:=(\pi_h,\pi_{h+1},...,\pi_n)\in \nn^{n-h}\times\zn$.
We define
${\cal H}_{\Pg}:=(p^n)^{\pi_{n-1}}...(p^{h+1})^{\pi_h}{\cal H}_{\pi_n,0}$.
Clearly the maps $p^{\pm l}:{\cal H}_{\Pg}\rightarrow
{\cal H}_{\Pg\pm e_{l-1}}$
are invertible ($p^{-l}$ only in the orthogonal complement of its kernel),
the inverse being
$[p^l]^{-1}=\f{q^{\rho_l}}{(p\cdot p)_l-(p\cdot p)_{l-1}}p^{-l}$
and $[p^{-l}]^{-1}=\f{q^{\rho_l}}{(p\cdot p)_l-q^{-2}(p\cdot p)_{l-1}}p^l$
respectively, as one can easily check using equations (119). Therefore we
arrive at the proposition

\begin{prop}
${\cal H}$ can be decomposed into the direct sum
\be
{\cal H}=\bigoplus\limits_{\Pg\in\nn^{n-h}\times\zn}{\cal H}_{\Pg},
{}~~~~~~~~~~~~~~~~
{\cal H}_{\Pg}:=\La^{\pi_n}(p^n)^{\pi_{n-1}}...(p^{h+1})^{\pi_h}
{\cal H}_{\vec{0}}
\ee
of orthogonal eigenspaces ${\cal H}_{\Pg}$ of the observables
$(p\cdot p)_i$,
\be
(p\cdot p)_l{\cal H}_{\Pg}=M^2q^{\sum\limits_{k=l}^n2(1+\pi_k)}{\cal H}
_{\Pg}~~~~~~~~~~~~~~~~~~l=h,h+1,...,n.
\ee
\end{prop}

In the case $N=2n+1$ we will attach superscripts $\pm$
when we want to specify  that we are dealing with irreps
characterized by positive (resp. negative) eigenvalues of $p_0$.
Then:
\be
p_0{\cal H}_{\Pg}^{\pm}=\pm M[1+q^{-1}]^{\f 12}
q^{\sum\limits_{k=0}^n(1+\pi_k)}{\cal H}^{\pm}_{\Pg}.
\ee

{\bf Remarks}. As expected the spectra of $(p\cdot p)_i$ are discrete;
they are particularly simple, since they consist only of q-powers.
Note that none of them contains the zero eigenvalue (but the latter is an
accumulation point of the spectra); in particular $(p\cdot p)_n>0$ always,
i.e. `` there is no state in which the nonrelativistic quantum particle
is at rest ''. Furthermore, note that fixing the values of
the observables $(p\cdot p)_i$ selects a $n$-Torus within momentum space,
in other words ${\cal H}_{\Pg}$ will consist of
states with a ``support '' confined in
$n$-Torus ${\cal T}_{\Pg}^n$ in momentum space.
As we will see, $h_i:=log_{q^2}\k^i$ ($i=1,2,...,n$) will play the
role of `` action variable ''
observables conjugated to the $n$ `` angle '' variables of the torus.

Assume that $|\phi>\in {\cal H}$ is an eigenvector
of $\k^i$: $\k^i|\phi>=\lambda_i|\phi>$ $\forall i=1,...,n$.
It is easy to
realize from eq.s (46),$(67)_1$
that application of all generators of $u_q(e^N)$ to $|\phi>$ will
yield eigenvectors of $\k^i$ with eigenvalues $\lambda_iq^{j_i}$,
with fixed $\lambda_i$ and
$\J:=(j_1,...,j_n)\in \zn^n$. We can assume without loss
of generality that $1\ge\lambda_i>q^2$.
We will see in the sequel which further restrictions  there are on
the values of $\lambda_i,j_i$.
Summing up, our Hilbert space will be spanned by orthonormal vectors
$|\Pg;\J;\alpha>$ such that
\be
(p\cdot
p)_l|\Pg;\J;\alpha>=M^2q^{\sum\limits_{k=l}^n2(1+\pi_k)}|\Pg;\J;\alpha>,
{}~~~~~~~~~~~~~~~~~~~~~~\k^i|\Pg;\J;\alpha>=\lambda_iq^{j_i}|\Pg;\J;\alpha>.
\ee
$\alpha$ stands for possible further labels necessary to completely identify
the vectors of a basis of ${\cal H}$. They occur if the set of
commuting observables (115) is not complete on ${\cal H}$; they label the
eigenvalues of the commuting observables which are to be added to the ones
reported in formula (115), to get a complete set.

In the case $N=2n+1$ we will attach superscripts $\pm$
when we want to specify  that we are dealing with irreps
characterized by positive (resp. negative) eigenvalues of $p_0$.
Then:
\be
p_0{\cal H}_{\Pg}^{\pm}=\pm M[1+q^{-1}]^{\f 12}
q^{\sum\limits_{k=0}^n(1+\pi_k)}{\cal H}^{\pm}_{\Pg}.
\ee

{}~~

In the sequel we look for a definition of the action of all the generators
of $u_q(e^N)$
on these vectors which is consistent with all the algebra relations;
maximal extension of the domain of these operators (in particular of the
symmetric ones, in order to get self-adjoint
operators in ${\cal H}$) is out of the scope of this work. To
analyze the action of the other generators of $\hat u_q(e^N)$
it is convenient to separate the action of changing
$\Pg$ from that of changing
$\J$ by using the operators $L$ instead of the operators $\M$;
in fact, relation (99) implies $L:{\cal H}_{\Pg}\rightarrow {\cal H}_{\Pg}$.
Then

\begin{theorem}
On a basis $\{|\Pg;\J,\alpha>\}$ ($\Pg\in\nn^{n-h}\times\zn$, $\J\in\zn^n$)
of ${\cal H}$
\be
p^l|\Pg;\J,\alpha>=M[1-q^{2(\pi_{l-1}+1)}]^{\f 12}q^{\sum\limits_{k=l}^n
(1+\pi_k)}|\Pg+\e_{l-1};\J+\y_l,\alpha'>
\ee
\be
p^{-l}|\Pg;\J,\alpha>=M[1-q^{2\pi_{l-1}}]^{\f 12}q^{-\rho_l+\sum\limits_{k=l}^n
(1+\pi_k)}|\Pg-\e_{l-1};\J-\y_l,\alpha'>.
\ee
Here $l>h$,$\e_l\in \nn^{n-h},\y_i\in \zn^n$
with $(\e_l)^j=\delta_l^j$, $(\y_i)^j=\delta^j_i$. Whereas
\be
\cases{ p_0|\Pg;\J,\alpha>=\pm M[1+q^{-1}]^{\f 12}
q^{\sum\limits_{k=0}^n(1+\pi_k)}|\Pg;\J,\alpha>,~~~~~~~~~~~if~~~N=2n\!+\!1,~~~
|\Pg;\J,\alpha>\in {\cal H}^{\pm}; \cr
p^{\pm 1}|\Pg;\J,\alpha>=Mq^{\sum\limits_{k=1}^n(1+\pi_k)}|\Pg;\J\pm y_1,
\alpha'>,~~~~~~~~~~~~~~~if~~~N=2n\cr}
\ee
(we have set all the arbitrary phase factors equal to 1). Moreover
\be
\cases{
L^{1-m,m}|\Pg;\J,\alpha>=D_m(\Pg,\J,\alpha)|\Pg;\J+y_m-y_{m-1},\alpha'> \cr
L^{-m,m-1}|\Pg;\J,\alpha>=D'_m(\Pg,\J,\alpha)|\Pg;\J-y_m+y_{m-1},\alpha'> \cr
\k^i|\Pg;\J;\alpha>=\lambda_1q^{2j_i}|\Pg;\J;\alpha>;\cr}
\ee
The coefficients $D_m,D'_m$ depend on the particular irrep under consideration;
the precise domain of $\J$ will be given in formula (152).
\end{theorem}
$Proof$. Let us consider for instance the proof of relation (126). One
starts from the ansatz $p^l|\Pg;\J,\alpha>=A|\Pg+\e_{l-1};\J+\y_l,\alpha>$,
takes the norm of this vector, uses hermitean conjugation and knowledge
of the eigenvalues of formula
(122) to find $|A|^2$; the arbitrary phase factor of $A$ is taken equal
to one for convenience. The equality $\lambda_i=\lambda_1\in[1,q^2)$ will be
proved in next subsection.
$\diamondsuit$

Formula $(128)_b$ implies that the range of $j_1$ in the case $N=2n$ is
$\zn$.

Theorem 4 summarizes the essential features of the promised
`` q-latticization '' in momentum space $\rn^N_{\vec{p}}$. For each vector
$\Pg$, the equations $(p\cdot p)_l=M^2q^{\sum\limits_{k=l}^n2(1+\pi_k)}$
single out a n-Torus  submanifold ${\cal T}^n$
within $\rn^N_{\vec{p}}$, on which the states
of ${\cal H}_{\Pg}$ have support. However, the additional specification
of a  vector $\J$ selects in ${\cal H}_{\Pg}$ state(s) having
well-defined angular momentum components $\k^i$, but no well-defined
$p$-angles; in other words the support of each state is $not$ concentrated
on a point of ${\cal T}^n\subset\rn^N_{\vec{p}}$.
For no choice of a complete set of commuting
observables the corresponding eigenvectors would have a point-like support
in $\rn^N_{\vec{p}}$, since no such set can include $N$ functions of the
(non-commuting variables) $p^i$'s. The q-lattice $\{(\Pg,\J)\}$ has to be
understood in a space where $n+1-h$ dimensions
(corresponding to the first $n+1-h$ observables (115))
 are of `` momentum '' type, and the remaining are  of
`` angular momentum '' type. The action of the generators $p,L,\La^{\pm 1}$
on a vector $|\Pg;\J,\alpha>$ can be visualized as a mapping of
the point $(\Pg;\J)$ of the q-lattice into one of its nearest
neighbour points.

In fig. 1 we draw the 1-tori ${\cal T}^1\subset \rn^3_{\vec{p}}$
(circles) and the corresponding action of $\La^{\pm 1},p^{\pm 1}$ in the case
$N=3$.

The number and the form of the extra paramaters $\alpha$, the domain of
$\J$'s, the values
of $\lambda_i$ of formula (124) and the explicit form of $D_m,D'_m$ can be
determined only after a detailed study of the structure of the subspaces
${\cal H}_{\Pg}$. Due to equations (98),(121), if we knew
the structure of any subspace ${\cal H}_{\Pg}$
and the way the $L$'s operators act on it, we would be able to extend
this knowledge in a straightforward way to all the other subspaces, through
application to ${\cal H}_{\Pg}$ of powers in the momenta.

\subsubsection{Structure of ${\cal H}_{\vec{0}}$}

As a particular subspace we take ${\cal H}_{\vec{0}}$; next,
we are going to investigate its structure.
Note that the `` central charges '' of formulae (103), (104) reduce to
\be
C_m|_{{\cal H}_{\vec{0}}}=\cases{0~~~~~~~~~~~if~~~~m>h+1 \cr
\f{q^{-\f 32}}{q^{-1}-1}~~~~~~~~~~~~~~if~~~~N=2n+1~~and~~m=1 \cr
\f{1}{1-q^2}~~~~~~~~~~~~~~~if~~~~N=2n~~and~~m=2 \cr}
\ee

Correspondingly the triples
${\cal T}_m:=L^{1-m,m},L^{-m,m-1},\k^m(\k^{m-1})^{-1}$
generate $U_q(su(2))$ subalgebras (since they satisfy the same
commutation relations of $M^{1-m,m},M^{-m,m-1},$
$\k^m(\k^{m-1})^{-1}$)
when $m>h+1$, whereas in the remaining cases they generate subalgebras
characterized by the following
modified relations:
\be
[L^{01},L^{-1,0}]_q=q^{\f 12}\f{q^{-1}+(\k^1)^{-1}}{1-q^2}
\ee
if $N=2n+1$ and
\be
\cases{[L^{1,2},L^{-2,1}]=\f{q^{-2}p^1p^1}{(1-q^2)(p\cdot p)_1},
{}~~~~~~~~~~
[L^{-1,2},L^{-2,-1}]=\f{q^{-2}p^{-1}p^{-1}}{(1-q^2)(p\cdot p)_1},~~\cr
[L^{-1,2},L^{1,2}]=0~~~~~~~~~~~~~~~~~~~~~~~~~[L^{-2,1},L^{-2,-1}]=0, \cr}
\ee
\be
[L^{\pm 1,2},L^{-2,\mp 1}]_{q^2}=\f{(\k^{2})^{-1}(\k^1)^{\mp 1}}{1-q^2},
\ee
if $N=2n$.

As a first task we want to determine the constants $\lambda_i$ involved
in the definition (124) of the eigenvalues of the $\k^i$.

To each triad ${\cal T}_m$, $m>h+1$, we can apply the representation
theory of $U_q(su(2))$.
${\cal H}_{\vec{0}}$ can be completely decomposed into the direct sum of the
representation spaces of the irreps of each $U_q(su(2))$
triple ${\cal T}_m$, separatly. Therefore well-known results
concerning the irreps of $U_q(su(2))$ imply that there exists
$l\in\nn$ characterizing each irrep of ${\cal T}_m$
such that the eigenvalues
of $log_q[(\k^m)(\k^{m-1})^{-1}]$ are $-l,1-l,...,l$, implying
$log_q(\f{\lambda_m}{\lambda_{m-1}})\in \zn$
in all ${\cal H}_{\vec{0}}$. We recall that these well-known results  follow
from the existence of both an highest and a lowest weight within each
irrep of $U_q(su(2))$.

It remains to evaluate $\lambda_1$. We are going to show that this
constant is not constrained
by the representation theory of the algebrae (131),[(132),(133)].

${\cal H}_{\vec{0}}$ can be completely decomposed into the direct sum of the
representation spaces ${\cal H}_{\vec{0}}^s$ of the $*$-irreps of these
latter algebras. Let us study them.

We start from the first algebra (i.e. with odd $N$).
It is immediate to realize that, because of the
embedding mentioned after Proposition 5,
the casimir $\Omega$ of the algebra (131) is formally given by
the casimir $\Omega=\Omega^1|_{{\cal H}_{\vec{0}}}$ of $\hat u_q(e^3)$, namely
(see formula (110))
\be
\Omega=L^{-10}L^{01}(\k^1)^{\f 12}+q^{\f 12}\f{(\k^1)^{-\f 12}-(\k)^{\f 12}}
{(q^2-1)(q-1)}.
\ee
Let $\Omega{\cal H}_{\vec{0}}^s=\omega^s{\cal H}_{\vec{0}}^s$,
$\omega^s\in\rn$, let
$|\psi>\in {\cal H}_{\vec{0}}^s$ be an eigenvector
of $\k^1$, $\k^1|\psi>=\mu^2|\psi>$, and define
$|\psi_m>:=(L^{01})^m|\psi>$. Then
\be
<\psi_{m+1}|\psi_{m+1}>=q^{-\f 32}<\psi_m|L^{-10}L^{01}|\psi_m>
=<\psi_m|\psi_m>\{\omega^sq^{-m}\mu^{-1}+q^{\f 12}\f{1-\mu^{-2}q^{-2m}}{
(q^2-1)(q-1)}\}.
\ee
Assuming $\mu>0$, we realize that $<\psi_{m+1}|\psi_{m+1}>$ would get negative
for large $m$ unless it vanishes for some $m$. The latter condition
means the existence of
a highest weight vector $|\tau_s>\in{\cal H}_{\vec{0}}^s$,
\be
L^{01}|\tau_s>=0,~~~~~~~~~~\k^1|\tau_s>=\tau_s^2|\tau_s>,
\ee
and $\omega^s=q^{\f 12}\f{\tau_s^{-1}-\tau_s}{(1-q^2)(1-q)}$. If we repeat the
same argument with $|\psi_{-m}>:=(L^{-10})^m|\psi>$, we see that its norm
keeps positive for large $m$, hence there exists no lowest weight
vector and no restriction on the value of $\tau_s$.
The initial vector $|\psi>$ can be reconstructed from $|\tau_s>$ through
$|\psi>=\alpha (L^{-10})^m|\tau_s>$ (with some $\alpha\in \rn$); in fact
each power of $L^{-10}L^{01}$ can be expressed as a
diagonal operator which is a
combination of $\omega^s(\k^1)^{-\f 12}$ and a function
of $\k^1$, therefore $(L^{-10})^m(L^{01})^m$ is diagonal.

If it were
$\mu<0$ we
would find a lowest weight vector and no highest weight, on the contrary;
for the sake of brevity, in the sequel we will assume that $\mu>0$..

Since, if $N=5,7,...$, different
subspaces ${\cal H}_{\vec{0}}^s\subset{\cal H}_{\vec{0}}$ are mapped into
each other by the remaining  $L$ operators -
as well as by the $p^i$ - in such a way that the $\k^1$
eigenvalues are only rescaled by even powers of q, we infer that the
constant $\lambda_1$ ($1\ge\lambda_1> q^2$) involved in the $\k^1$ eigenvalues
is characteristic of the $\hat u_q(e^N)$ irrep and unconstrained.
Its value is a function of the casimirs of the irrep.
Note that no classical analogue of such representations with
$1>\lambda_1>q^2$, $\lambda\neq q$
is available.

Now we consider the $N=2n$ case. Since the expression for the casimir
of the algebra (132)+(133)
is not so simple, we prefer to use the Lemma 3 of the appendix
to prove the existence of highest weight vectors (in the sense of vectors that
are annihilated by $L^{\pm 1,2}$). In fact there are an infinite series, for,
if $|\phi>$ is one, then $(\f{p^{\pm 1}}M)^l|\phi>$ is an independent one
$\forall l\in \zn$. For proving that the whole representation space can
be reconstructed from the highest weight vectors one needs using the
explicit expression for the casimir, and we omit the computations here.
Reasoning as in the $N=2n+1$ case one concludes as before that there is a
unique
constant $\lambda_1$ (involved in the $\k^1$ eigenvalue) characterizing the
irrep.

{}~

Let $P^{+,N}$ be the subalgebra of $\hat u_q(e^N)$ generated
by $p^l$, $l>h(N)$, and let
\be
u_q^{c,N}:=u_q^{-,N}\ot P^{+,N}\ot \cn[\La,\La^{-1}]
\ee
($u_q^{\pm,N}$ were defined at the end of section III.3).
As a second point comes
\begin{theorem}
The subspace ${\cal H}_G$ of `` highest weight vectors '', i.e.
\be
{\cal H}_G:=\{|\phi>\in {\cal H}|_{\pi_n=0},~~~~L|\phi>={\bf 0},~~~~~~
p^{-l}|\phi>={\bf 0}~~~~~~~~~~\forall L\in u_q^{+,N},~l>h \}
\ee
is nonempty and ${\cal H}_{\vec{0}}=u_q^{-,N}{\cal H}_G$.
It is one-dimensional in the
case $N=2n+1$ and infinite dimensional in the case $N=2n$. In the latter
case a basis of ${\cal H}_G$ is provided by the vectors
$\{(p^{\pm 1})^r|\phi>,~~r\in \nn$\}, where $|\phi>$ is any nontrivial
vector of ${\cal H}_G$. Using the results of Proposition 8 we can say that
$|\phi>$ is cyclic in ${\cal H}$ w.r.t. the subalgebra $u_q^{c,N}$.
(In the sequel by `` the highest weight vector '' we will mean
a particular one of these vectors, in the case $N=2n$).
The eigenvalues $k^i$ of the operators $\k^i$ are of the type
$k^i=q^{2j_i}\lambda_1$, $j_i\in\zn$, and the constant $\lambda_1$,
$1\ge\lambda_1>q^2$, is a function of the casimirs characterizing the irrep.
\end{theorem}
$Proof$. We note
that it is sufficient to prove the theorem in ${\cal H}_{\vec{0}}$, by
proposition 41. The proof for $N=3,4$ amounts to the discussion preceding
the claim. The~general proof will be partially  given in the appendix.

{\bf Remark:} The existence of highest weight vectors follows from the
requirement $(114)_1$. Contrary to the case of representation theory of
$U_q(so(N))$, there is no lowest weight vector in ${\cal H}_{\vec{0}}$, due
to the presence
of non-vanishing $C_1,C_2$ in the commutation relations (131), (133) when
$N$ is respectively odd, even.
For this reason the application
of $u_q^{-,N}$ to a highest weight vector generates
an infinite-dimensional space (the whole ${\cal H}_{\vec{0}}$ when $N=2n+1$,
its subspace characterized by odd/even $j_1$ if $N=2n$).

{}~~~

In the sequel we will stick to irreps
characterized by $\lambda=1,q$ (among which we can find those having classical
analogue).

For this class of irreps we can introduce a vector $\w\in \zn^n$ such that
$\k^i|\phi>=q^{w_i}|\phi>$.
The vector $\w$ depends on the casimirs and together with the value of $M$
completely characterizes an irrep.
We will therefore attach it as a superscript to the symbol ${\cal H}$
and write ${\cal H}^{\w}$ when we want to specify that we are considering the
irrep with highest weight $\w$; correspondingly, we
will attach superscripts to the ket symbols: $|\vec{0},\J,...>^{\w}$;
the highest weight vector itself will be denoted by $|\vec{0},\w>^{\w}$.
So far we have avoided an heavy notation distinguishing the abstract generators
$p,L,\k,\M$ of the Euclidean algebra from their realizations as
operators belonging to a particular representation of the algebra.
{}From now on we will denote by $\Gamma^{\w}$ the Irrep with highest weight
$\w$, and we will write $\Gamma^{\w}(p),\Gamma^{\w}(L), ...$  instead of
$p,L,...$ when we want to stress that we are dealing with operators
on ${\cal H}^{\w}$ representing $p,L,...$.

{\bf Definition} We define the
singlet Irrep as the one characterized by the highest weight
$\w=0$. It will play a crucial role in the sequel, as one expects from
comparison with the classical case.

It is immediate to verify that on the singlet Irreps of
$\hat u_q(e^3),~\hat u_q(e^4)$
the casimirs $\Omega_1$ take zero values (see formulae (110),(111)),
since they annihilate
the corresponding highest weight states. This is no surprise, since they
vanish in the classical case as well.

{}~~

Now we are going to determine possible highest weights $\w$ and construct
generic (not necessarily singlet) Irrep
of $\hat u_q(e^N)$ from tensor products.
It is easy to verify that by making the tensor product of the singlet
Irrep $(\Gamma^{\vec{0}},{\cal H}^{\vec{0}})$ of $\hat u_q(e^N)$
and an Irrep $(\Gamma^{\vec{u}}_{hom},{\cal H}^{\vec u}_{hom})$
of $U_q^N\equiv U_{q^{-1}}(so(N))$ with highest weight $\vec{u}$  we find
a reducible Hilbert space  representation of $\hat u_q(e^N)$ characterized
by the same mass, as it occurs
in the classical case. In fact, let
$\tilde {\cal H}^{\vec{u}}:={\cal H}^{\vec{0}}\ot {\cal H}^{\vec u}_{hom}$
and define $\tilde \Gamma^{\vec{u}}$ on $\tilde {\cal H}^{\vec{u}}$ by
\be
\tilde \Gamma^{\vec{u}}:=(\Gamma^{\vec{0}}\ot \Gamma^{\vec{u}}_{hom})
\circ \phi,~~~~~~~~~~~~~~~~~\Gamma^{\vec{u}}_{hom}(p^i):=0=:\Gamma^{\vec{u}}_
{hom}((p\cdot p)_i),~~~~~~~~\Gamma^{\vec{u}}_{hom}(L):=\Gamma^{\vec{u}}_{hom}
(\M),
\ee
where $\phi$ is the coproduct of $U_q(e^N)$. Then it is immediate to verify
that $\tilde \Gamma^{\vec{u}}(p^i),\tilde \Gamma^{\vec{u}}(L^{ij}),
\tilde \Gamma^{\vec{u}}(\k^i)$ satisfy the commutations of $U_q(e^N)$
and the spectra of the operators $\tilde \Gamma^{\vec{u}}((p \cdot p)_i)$
are the same as in the
singlet Irrep (i.e. the `` mass '' scale $M$ characterizing the Irrep
is the same); in fact, for instance, one can immediately verify that
$$
[\tilde \Gamma^{\vec{u}}(L^{-m,m+1}),\tilde \Gamma^{\vec{u}}(L^{-m-1,m})]_{q^2}
=q^{1+2\rho_m}\f{1\ot 1 -\k^(\k^{m+1})^{-1}\ot \k^m(\k^{m+1})^{-1}}
{q-q^{-1}}+C_m\ot 1
$$
\be
=\tilde \Gamma^{\vec{u}}\left[\f{1-\k^m(\k^{m+1})^{-1}}{q^2-1}q^{2\rho_m}+C_m
\right ]~~~~~~~~~~~m\ge 1
\ee
where the central charges $C_m$ were defined in formulae (103),(104).

We would like now to single out the Irreps contained in
$\tilde \Gamma^{\vec{u}}$.
In the classical theory this can be done imposing
the `` wave-equations '' on the tensor product space; each kind of wave
equation
selects the subspace corresponding to an Irrep, which consists of the tensor
product of a one-dimensional representation (characterized by a vector
$\vec{p}$) of the translation subalgebra and an Irrep of the little
subgroup $SO(N-1)\subset SO(N)$
of the direction of $\vec{p}$ in the momentum space. An equivalent approach
is to tensor this one-dimensional representation directly to the little
group of $\vec{p}$.  It seems difficult to apply either approach
in the q-deformed case, since we don't
know natural embeddings $U_q(so(N-1)) \hookrightarrow U_q(so(N))$,
except when $N=3,4$. When $N=3$ we have in fact a natural embedding
$U(so(2))\ap U(1) \hookrightarrow U_q(so(3))$: $U(so(2))$ is the classical
subalgebra generated by $\k^1$. In the case $N=4$ we have also a
natural embedding, since
$U_q(so(3)) \ap U_q(su(2))\hookrightarrow U_q(su(2))\ot U_q(su(2))\ap
U_q(so(4))$

However, there is a natural way to find for any $N$
the Irreps contained in $\tilde \Gamma^{\vec{u}}$, due to the
fact that the representation is of highest weight type; it is the
usual procedure that e.g. one uses to determine
the irrep decomposition of the tensor product of two irreps of the same
(classical) Lie algebra. Using orthogonality, one determines all the
highest weight vectors contained in ${\cal H}^{\vec{u}}$.

For the sake of being explicit let us consider the case $N=2n+1$. The vector
\be
|\vec{0},\vec{u}>^{\vec{u}}:=|\vec{0},\vec{0}>^{\vec{0}}\ot ||\vec{u}>
\ee
is clearly an highest weight vector of $u_q(e^N)$ (in the sense of theorem 1)
and when we apply $u_q^{-,N}$ to it we generate the Irrep
$(\Gamma^{\vec{u}},{\cal H}^{\vec{u}})$; in particular
$\Gamma^{\vec{u}}(L^{-1,0})|\vec{0},\vec{u}>^{\vec{u}}\in {\cal H}^{\vec{u}}
\subset \tilde{\cal H}^{\vec{u}}$.

Since the subspace of $\tilde{\cal H}^{\vec{u}}$ with $\J=\vec{u}-\y_1$
is two-dimensional (it is spanned by the vectors
$|\vec{0},\vec{0}>^{\vec{0}}\ot ||\vec{u}-\y_1>$,
$|\vec{0},-\y_1>^{\vec{0}}\ot ||\vec{u}>$), its orthogonal complement
is one-dimensional and we can easily verify that it is spanned by
\be
|\vec{0},\vec{u}-\y_1>^{\vec{u}-\y_1}:=
[q^{-\f 32}(u_1)_{q^{-2}}]^{\f 12}|\vec{0},-\y_1>^{\vec{0}}\ot ||\vec{u}>
-\left[\f{q^{\f 52}(1+\lambda_1q)}{q^{-2}-1}\right] ^{\f 12}
|\vec{0},\vec{0}>^{\vec{0}}\ot ||\vec{u}-\y_1>
\ee
(as usual, the tensor product scalar product is defined on the
vector of a basis
as the product of the scalar product of the tensor factors and is extended
by linearity).

$|\vec{0},\vec{u}-\y_1>^{\vec{u}-\y_1}$ clearly is
a highest weight vector itself, because $\tilde {\Gamma^{\vec{u}}}$ is
a $*$-representation, and when we apply $u_q^{-,N}$ to it we
generate a different Irrep,
$(\Gamma^{\vec{u}-\y_1},{\cal H}^{\vec{u}-\y_1})$. It is evident that
if we reiterate this procedure $2u_1$ times we find $2u_1+1$ Irreps.

The same argument can be applied in the case $N=2n$ to the highest
weight vector of $(\tilde \Gamma^{\vec{u}},\tilde {\cal H}^{\vec{u}})$.
We conjecture that all highest weights (for the class of representations
with $\lambda_1=1,q$) can be obtained in this way.
The final result is summarized in the

\begin{prop}
The irreps of $u_q(e^N)$ (characterized by $\lambda_1=1,q$)
are highest weight irreps. Possible highest weights are
of the form $\Pg\equiv 0$, $\w\equiv\vec{u}-l\y_1$, $0\le l\le 2u_1$,
if $N=2n+1$,
$\w\equiv\w(l,l'):=\vec{u}-l\cdot sign(u_2-u_1)(\y_2-\y_1)-l'(\y_2+\y_1)$,
$0\le l\le |u_2-u_1|$,  $0\le l'\le u_1+u_2$, if $N=2n$; $\vec{u}$
denote weights of $U_q^N$ . In particular, when $N=3,4$
the sets $\{\w\}$ of weight satisfy the relations $\{w_1\}=\zn$,
$\{\w\}\subset \zn\ot \zn$ respectively.
We have the following tensor product decomposition
\be
\tilde {\Gamma}^{\vec{u}}=\cases{\bigoplus\limits_{l=0}^{2u_1}
\Gamma^{\vec{u}-l\y_1}~~~~~~~~if~~~N=2n+1 \cr
\bigoplus\limits_{0\le l\le |u_2-u_1|; \atop 0\le l'\le u_1+u_2}
\Gamma^{\w(l,l')}~~~~~~~~if~~~N=2n+1 \cr}
\ee
Highest weight vectors can be easily determined from the above described
tensor product construction procedure.
\end{prop}

Note that the irreps with $w_1<0$ have no classical analogue.

\subsubsection{Moding out singular vectors in the singlet irrep}

{}~~~According to theorem 5,
${\cal H}_{\vec{0}}$ is generated by application of
the Borel subalgebra $u_q^{-,N}$ to its highest weight vector.
A Poincar\'e-Birkhoff-Witt basis for $u_q^{-,N}$ is the set of monomials
$(L_{\alpha_1})^{m_1}...(L_{\alpha_s})^{m_s}(p^{sign(m)\cdot 1})^{|m|}$,
where ~~
$s=\cases{n^2~~~if~~N=2n+1 \cr n(n-1)~~~if~~N=2n\cr}$;
 by $L_{\alpha_i}$
we denote the $L^{j,k}$ corresponding to the negative root $\alpha_i$
($i=1,2,...,s$), and
the roots have been ordered  according to an admissible total order
($\alpha_i>\alpha_l$ if $i>l$). Then, to each such monomial there
corresponds a vector of ${\cal H}_{\vec{0}}$; not all of these vectors,
however, can be considerated as independent.

In fact, as in the classical case, the requirement that the
representation is of Hilbert space type makes many combinations
of the previous
vectors singular. We remind that a vector $|\chi>$ is said to be singular if
$<\psi|\chi>=0$ $\forall |\psi>\in {\cal H}_{\vec{0}}$.
The simplest examples of singular vectors can be found
in the singlet representation when $N>4$.
They are $L^{-i,i-1}|\vec{0},\vec{0}>$, where $i>h+1$. Actually they are
orthogonal to all vectors with different weights, by the hermicity of the
$\k^l$, and have zero norm because of eq $(102)_1$, $(130)$ and the definition
of the singlet irrep. Therefore we have to set
\be
L^{-i,i-1}|\vec{0},\vec{0}>=0
\ee
if we want the basic requirement $(114)_2$ of a Hilbert space to be satisfied.
A slightly more complicated example is provided in the same representation
and for $N\ge 5$ odd by the vectors
$|1>:=L^{-1,0}L^{-2,1}L^{-1,0}|\vec{0},\vec{0}>$,
$|2>:=L^{-2,1}(L^{-1,0})^2|\vec{0},\vec{0}>$,
since we find, after straightforward use of formula $(102)_1$, Proposition 5,
\be
|2'>:=|1>-\f 1{q+q^{-1}}~~~~~~~~\Rightarrow~~~~~~~~~
<1|2'>=0=<2|2'>~~~~~~~~\Rightarrow~~~~~~~~~~<\psi|2'>=0~~~~\forall \psi
\in {\cal H}_{\vec{0}}.
\ee
When $N=2n$ another example of singular vector is
\be
|\tau>:=(L^{-2,-1}-q^2L^{-2,1}\f{p^{-1}p^{-1}(\k^1)^{-1}}{(p\cdot p)_1})
|\vec{0},\vec{0}>,
\ee
as it can be easily verified by applying commutation relations (100),$(102)_1$.

An Irrep of our algebra is obtained by moding out
all singular vectors which are generated in the previous construction.
Now we want to identify them.

Let us stick for the moment to the  singlet  representation of $\hat u_q^N$.

\begin{theorem}
In the singlet Irrep the commutation relations
\be
\cases{
g_1:=L^{-2,-1}-q^2L^{-2,1}\f{p^{-1}p^{-1}(\k^1)^{-1}}{(p\cdot p)_1}=0\cr
\tilde g_1:=L^{-2,1}-q^2L^{-2,-1}\f{p^1p^1\k^1}{(p\cdot p)_1}=0\cr}
{}~~~~~~~~~~~~~~~~if~~~N=2n,
\ee
\be
g_i:=L^{-i,i-1}L^{-i-1,i}-\f{(\k^i)^{-\f 12}-(\k^i)^{\f 12}}
{q(\k^i)^{-\f 12}-q^{-1}(\k^i)^{\f 12}}L^{-i-1,i}L^{-i,i-1}=0
{}~~~~~~~~~~~~i> h
\ee
and their hermitean conjugates hold.
All singular vectors of the singlet representation can be obtained by
application either of any element of $\hat u_q^{-,N}$ to a singular vector,
or of some $g_i$ to a nonsingular vector.
Therefore if we mode out the singular
vectors, an orthonormal basis ${\cal B}_q$
of ${\cal H}_{\vec{0}}$ will consist of the vectors
\be
|\vec{0},-\J>:\propto(L^{-n,n-1})^{J_n}(L^{1-n,n-2})^{J_{n-1}}...
(L^{-2,1})^{J_2}\cdot
\cases{(L^{-1,0})^{J_1}|\vec{0},\vec{0}>~~~~~if~~N=2n+1 \cr
(p^{- sign(J_1)\cdot 1})^{|J_1|}|\vec{0},\vec{0}> ~~~~~if~~N=2n \cr}
\ee
where the integers $J_i$ satisfy the conditions
$J_{i+1}\le J_i~~~~if~~i\ge h+1.$
The integers $\J\in \nn^n$ specify the eigenvalues of $\k^i$
and are given by
\be
j_i=J_i-J_{i+1}\ge 0~~~~~~~~~~~~~~~~~~~~~~~(J_{n+1}\equiv 0)
\ee
\end{theorem}
$Proof$: see the appendix.

{\bf Remarks}
\begin{itemize}
\item It is easy to show that the above relations are compatible
with the Serre relations of Proposition 5, and the commutation relations (98),
namely they yield identities $0=0$ when
plugged into the latter. The validity of commutation relations (147),
(148) depends  crucially
on the fact that $C_1\neq 0,~C_2\neq 0$ in relations (102).

\item As a consequence of formula (149), each weight in the singlet
irrep has multiplicity 1, in other words we don't need extra parameters
$\alpha$ to identify the vectors of the basis ${\cal B}_q$.
This is a strong indication that it is
convenient to introduce a `` configuration space
realization '' of the singlet
irrep; actually we will see in section IV.2 how to represent its
vectors as functions on the quantum Euclidean space.
\end{itemize}

By inverting relations (150) one gets the equivalent relation:
\be
J_i=\sum\limits_{l=i}^n j_i
\ee

{}~

Let us consider now a generic representation $(\Gamma^{\w},{\cal H}^{\w})$.
Its singular vectors can be determined using relations (147),(148) in
${\cal H}^{\vec{0}}$ and the tensor product construction of
$\Gamma^{\w},{\cal H}^{\w}$, described in Proposition 9.
Of course, in building ${\cal H}^{\w}$ through application of
$u_q^{c,N}$
to the highest weight vector $|\vec{0},\w>^{\w}$, one has to impose
relations (147),(148) only on the generators $L$ acting on the first tensor
factor (the singlet representation) of the product (139).
The existence of singular vectors in ${\cal H}^{\w}$
is therefore only due to the existence of singular vectors in the
representation $\Gamma^{\w}$ of $U_q^N$.

\subsubsection{Summary: the structure of ${\cal H}^{\w},{\cal H}^{\vec{0}}$}

Finally, let us summarize and comment a little on the structure of the
pre-Hilbert spaces ${\cal H}^{\w}$.

First, we note that the
domain of the labels $\J$ of the vectors $|\Pg,\J,\alpha>$,
$\k^i|\Pg,\J,\alpha>=q^{2j_i}|\Pg,\J,\alpha>$, forming an
orthogonal basis of ${\cal H}^{\w}_{\Pg}$ is:
\be
{\cal J}:=\{\J\in \zn^n~~|~~j_i\le \pi_{i-1}+w_i,~~~~i=h+1,h+2,...,n;
{}~~~~~~j_1\in\zn~~~~if~~N=2n.\}
\ee
This follows from inequalities (150), the tensor product construction of
Proposition 9 and formulae $(98)$.

{\bf Remark}. The explicit form of the highest weight vectors (which can be
determined according to the procedure sketched in the proof of Proposition 9)
and the knowledge of the singular vectors of the singlet representation allow
in principle to fix for each $\w$ the extra parameter $\alpha$ appearing in
formulae (26),(129) and the coefficients $D_m,D'_m$ appearing
in Theorem 4.

We give an intuitive picture of the physical content of the spectra of the
observables (115) in the singlet representation. The subspace
${\cal H}_i^{\vec{0}}:=\bigoplus\limits_{\{\Pg,~|~\pi_{i-1}=0\}}
{\cal H}_{\Pg}^{\vec{0}}$ is the eigenspace of the observable
$p^{-i} p_{-i}=(p \cdot p)_i-(p \cdot p)_{i-1}$ with the minimum
eigenvalue compatible with a given eigenvalue of $(p \cdot p)_i$,
namely $p^{-i} p_{-i}=M^2q^{\sum\limits_{k=i}^n2(1+\pi_k)}(q^2-1)$;
the latter quantity never vanishes when $q\neq 1$. This means that the there is
always a
`` point zero '' momentum component available in the plane of the
coordinates $i,-i$.
Now let us ask in which `` directions '' of this plane this point zero
momentum component can be pointed.

The admitted eigenvalues of $ln_q(\k^i)$,
i.e. of the angular momentum component in the plane, are $j_i\le 0$
(see (150)) and show that (except when $N=2n$, $i=1$)
only a `` clockwise '' or `` radial '' orientation are possible
The anticlockwise is excluded!
If $N=3$, for instance, minimum $p^1p_1$ means that the momentum is
`` almost pointed '' in the $p^0$ direction; $j_1$ represents the
$p^0$-direction component of the (orbital) angular momentum. We find a
sort of a purely `` kinematical '' PT (parity+ time inversion) asymmetry of the
allowed momentum
space (under PT $p$ would remain unchanged, whereas $j_1$ changes sign), which
is a surprising feature for a lattice theory; in fact, at least
usual equispatiated lattice theories, which are commonly used
nowadays for regularization purposes, cannot have a parity
asimmetry by a well-known no-go-theorem \cite{nie}.
In next section we will see in which sense
in the classical limit $q\rightarrow 1$, however, parity
symmetry is recovered.

Both for odd and even $N$ the
value of $j_i$ is not bounded from below; larger and larger absolute values
of the angular momentum  with a fixed  amount of momentum available
in the plane $i,-i$ (in particular, with the
minimum amount $p^{-i} p_{-i}=(p \cdot p)_i-(p \cdot p)_{i-1}$) can be
intuitively described only if the
corresponding states have larger and larger values of the mean distance from
the origin of the plane. Thus we can visualize the states of
${\cal H}_i^{\vec{0}}$ with larger and larger $|j_i|$ as states with larger
and larger mean distance from the origin in the $x^i,x^{-i}$ plane.

\subsection{Configuration space realization of the singlet irrep}

{}~~~As known, the undeformed
Euclidean algebra $e^N$ can be realized
in the singlet representation
as the algebra of differential operators
acting on suitable subspaces of the algebra of `` functions ''
on $\rn^N$ configuration space
(e.g. the subspace of smooth functions, or square integrable, or
distributions...).
Working in configuration-space
realization rather than at an abstract level is very useful
in the classical context for many purposes;
for instance,  questions regarding in concrete cases
the domain of definition of some elements
of $U(e^N)$, considered as operators in the singlet representation,
- e.g. the essential self-adjointness domain of would-be observables,
Fourier (anti)transforms performed by integration in
configuration (or momentum) space, etc - are
best treated in configuration (or, depending on cases, momentum) space
realization. The occurence of distributions is
(ultimately) related to the the fact that the spectrum of each
position/momentum operator is continuous.

In the case $q\neq 1$ the spectrum is discrete, as we have seen, and the
corresponding eigenvectors are normalizable. In the
preceding sections we have investigated in an abstract way the singlet
irrep of $u_q(e^N)$, we would now be happy to find its corresponding
q-deformed
configuration-space counterpart, if any, in view of further developments
of the theory (concerning, for instance, domain or Fourier transform
questions).
The existence of such a counterpart is expected also because
theorem 6 shows that the notation with $L$ generators applied to
$|\vec{0},\vec{0}>$ is too heavy to identify a basis of ${\cal H}^{\vec{0}}$
once one has moded out the singular
vectors. The vectors of the basis ${\cal B}_q$ of formula (149)
are labeled just
by the corresponding powers $\pi_m,J_l$ of the generators of
$u_q^{c,N}$ applied on $|\vec{0},\vec{0}>$;
in other words these vectors don't depend (apart from a factor)
on the order in which these generators are applied. This
result can be obtained in a natural way in a configuration-space realization.

The standard way to represent momenta operators $p^i$ satisfying the reality
relations (84) through q-derivatives would be to define them as the real
combination (85); then, the scalar product of two vectors would be computed
formally as in the case of undeformed quantum mechanics by a q-integral
involving the corresponding two wave-functions.
Concrete computations in this case would be however
extremely hard; in fact,
the derivation relations of $p^i=q^N\p^i+\bp^i$ with $x$ have no simple
expression such as (8) in terms of some $\hat R$ matrix, but involve
$\La,l^{ij},B$ in the RHS, so that solving
the eigenvalue equations gets rather involved.
In Ref. \cite{fio2} we solved a similar problem in studying another quantum
mechanical physical system (the harmonic oscillator) on $\rn_q^N$
by using a non-standard way of realizing the momentum
operators $p^i$ in `` configuration space representation '' as a $pair$
of realizations such as $\p,\bp$.
We adopt the same approach here (see also Ref. \cite{fio5}).
As for the generators $\M,\k$ of the subalgebra $U_q^N\subset U_q(e^N)$, the
choice of the standard or nonstandard approach is irrelevant, since they
are represented by the same differential operators in both of them
and therefore their hermitean conjugation will amount to
$*$-conjugation in $Dif\!f(\rn_q^N)$.

We can introduce a $pair$ of realizations
(what we call the unbarred and the barred) of the algebra of
$U_q(e^N)$, i.e. a pair of algebra homomorphisms $\rho,\br$
\be
\begin{array}
{ccccccccc}
                       &      &  Dif\!f(\rn_q^N),~~Fun(\rn_q^N) & \\
                       & \nea &                                         & \\
U_q(e^N),~~{\cal H}    &      &                                         & \\
                       & \sea &                                         & \\
                       &      &  Dif\!f(\rn_q^N),~~Fun(\rn_q^N),  &
\end{array}
\ee
through
\be
\cases{\rho(p^i):=-i\La^a\p^i,\cr \br(p^i):=-i\bp^i\La^{-a}q^{-N(1+a)}\cr},
{}~~~~~~~~~~~~\rho(u):=u=:\br(u)~~~\forall u\in U_q^N~~~~~~~~~~~~~
\rho({\bf \La}):=\La:=\br({\bf \La})
\ee
($a\in \zn$).
It is easy to check that the hermitean conjugation in
$u_q(e^N)$, defined in formulae (4.30),(4.90) (we call it $\dagger$ here, to
avoid confusion) satisfies the
relation
\be
\rho(u^{\dagger})=[\br(u)]^*~~~~~~~~~~~~~~~~~~
\br(u^{\dagger})=[\rho(u)]^*,
\ee
where $*$ denotes the complex conjugation in $Dif\!f(\rn_q^N)$.
To be precise we define the homomorphisms $\rho,\br$
first on the vectors of the basis ${\cal B}_q$
(see theorem 6), and then we extend them linearly to all of ${\cal H}$.

The $\rho,\br$-images of the vectors
of the basis ${\cal B}_q$ are determined  by the requirement
that they satisfy q-differential equations which are
respectively the $\rho$- and $\br$-image of the equations
satisfied by $|\phi>$. Actually we can limit ourselves to the search of
$\rho$- and $\br$-images of the highest weight vector(s), since they are
cyclic in the representations spaces.
If we find solutions to these
differential equations we obtain (two) representations of $u_q(e^N)$;
then we have to see whether they are $*$ representations and, if so,
identify them with some irreps studied in the previous section.

This is the program before us.
We will see that it can carried through, and that we obtain
a $unique$ configuration-space realization of
the singlet irrep based on the use of a $pair$ of non-$*$-representations
(the unbarred and the barred).

All choices of $a$ in equation (154) are essentially equivalent.
The choice $a=-1$ will be particularly convenient for representing
scalar products in ${\cal H}^{\vec{0}}$ as integrals
of functions on $\rn_q^N$, and we adopt it here.

Then we can easily prove the important
\begin{lemma}
If the vector $|\phi>$ satisfies the equations
\be
(A+A^ip_i+u)|\phi>=0,~~~~~~A\in\cn,~u\in U_q^N,~~~~~~~~~
{}~~~~~~~~~~(p\cdot p)_n|\phi>=M^2|\phi>
\ee
and $\varphi:=\rho(|\phi>)\in Fun(\rn_q^N)$ is the function that
represents $|\phi>$ in the unbarred realization with $a=-1$, then its
barred partner $\bar{\varphi}:=\br(|\phi>)$ is given by
\be
\bar{\varphi}(x)=e_{q^2}[-m^2x\cdot x]\varphi(q^{-1}x)~~~~~~~~~~~
m^2:=(q^{-2}-1)q^{-2\rho_n-1}M^2
\ee
\end{lemma}
$Proof.$ After a shift $x\rightarrow q^{-1}x$, $\p\rightarrow q\p$
the hypothesis reads
\be
(A-iA^i\La^{-1}q\p_i+u)\varphi(q^{-1}x)=0,~~~~~~a\in\cn,~u\in U_q^N,~~~~~
{}~~~~~~~-q\La^{-2}(\p\cdot \p)_n\varphi(q^{-1}x)=M^2\varphi(q^{-1}x);
\ee
then the thesis can be verified by a straightforward calculation, using the
expression (22) of $\bp$ in terms of $\p,x$ and formulae (18),(19).
$\diamondsuit$

Let us define
\be
e_q[Z] :=\sum\limits_{n=0}^{\infty}{Z^n \over (n)_q!},~~~~~~~~
\varphi^J(Z):=\sum\limits_{n=0}^{\infty}\f{[-Z]^n}
{(n)_{q^{-2}}!(n+J)_{q^{-2}}!},~~~~~~~~~~~~
(n)_q:=\f{q^n-1}{q-1}.
\ee
Now comes the

\begin{prop}
One can solve the highest weight vector conditions (138) in the $\rho,\br$
realization. The resulting representations of
$\hat u_q(e^N)$ on $Fun(\rn_q^N)$ coincide with the singlet
representation $(\Gamma^{\vec{0}},{\cal H}^{\vec{0}})$.
The cyclic highest weight
vector $|\vec{0},\vec{0}>^{\vec{0}}$ is represented in the unbarred
realization by
\be
\rho(|\vec{0},\vec{0}>^{\vec{0}})
=:\varphi_0=\cases{e_{q^{-1}}[iM_0x^0],~~~~~~~~M_0=\pm(1+q^{-1})^{\f 12}q^{n+1}
{}~~~~~~~~~~if~~~N=2n+1 \cr
\sum\limits_{n=0}^{\infty}\f{[-(x\cdot x)_1M_0^2]^n}{[(n)_{q^{-2}}!]^2}
{}~~~~~~~~M_0=q^{n+\f{1}{2}}M~~~~~~~~~~~~~if~~~N=2n. \cr}
\ee
The subspace $\rho({\cal H}_{\vec{0}}^{\vec{0}})\subset \rho({\cal
H}^{\vec{0}})$
is spanned in the unbarred representation by $\rho({\cal B}_q)$, whose
elements are the functions
\be
\rho(|\vec{0},-\J>^{\vec{0}})\propto (x^{-n})^{j_n}...(x^{-2})^{j_2}\cdot
\cases{(x^{-1})^{j_1}e_{q^{-1}}[iM_0x^0] ~~~~~~~~if~~~N=2n+1 \cr
(x^{-sign(j_1)\cdot 1})^{|j_1|}\varphi^{J_1}((x\cdot x)_1M_0^2)
{}~~~~~~~~if~~~N=2n \cr}
\ee
where $J_1$ was defined in formula (151).
\end{prop}
$Proof$. The most general expression for the function
representing the cyclic vector is
\be
\varphi_0=\sum\limits_{l=-n}^n\sum\limits_{i_l=0}^{\infty}A_{i_{-n},...,i_n}
(x^{-n})^{i_{-n}}...(x^n)^{i_n}.
\ee
The requirement that it is annihilated by $\rho(p^{-i})$, $i>h$, implies that
there can be no dependence on $x^i$ (take in the order $i=n,n-1,...$ and
perform the derivations).
Similarly, since on
${\cal H}_{\vec{0}}$ $L^{1-i,i}=\M^{1-i,i}+ cost\cdot p^{i}p^{1-i}$,
the requirement that it is annihilated by $L^{1-i,i}$ implies that there
can be no dependence on $x^{-i}$. This is straightforward to check
when $i>h+1$, and a little more lenghty when $i=h+1$. In the case $N=2n+1$,
for instance, it is easy to check that
\be
L^{0,1}\varphi_0=0~~~~~\Rightarrow~~~~~~\cases{
A_{i_{-1},i_0+1}\propto \f{A_{i_{-1},i_0}}{(i_0)_{q^{-1}}}
{}~~~~~~~~~~when~~i_{-1}\ge 0 \cr
A_{i_{-1},i_0+1}\propto \f{A_{i_{-1},i_0}}{(i_0+1)_{q^{-1}}}
{}~~~~~~~~~~when~~i_{-1}\ge 1; \cr}
\ee
the first (resp. second) condition comes from setting the coefficient of
$(x^{-1})^{i_{-1}}(x^0)^{i_0-1}x^1$ (resp.
$(x^{-1})^{i_{-1}-1}(x^0)^{i_0+1}$) equal to zero. They are incompatible,
therefore $A_{i_{-1}i_0}=0$ if $i_{-1}>0$.
Finally the requirement that $\varphi_0$ is an eigenvector of
$\rho(p^0)$ in the case $N=2n+1$ or $\rho((p\cdot p)_1),\k^1$ in the case
$N=2n$ with the prescribed eigenvalues (following from the conditions
$(p\cdot p)_n=M^2$, $p^{-l}|\phi_0>=0$) yields the expression (160).

A direct application of the commutation relations (55),(57) yields
the expression in formula (161) as the basis
of $\rho({\cal H}_{\vec{0}}^{\vec{0}})$;
the commutation relations (147), (148) can be easily checked applying them
to the elements of this basis.

Thus we have obtained two representations of $u_q(e^N)$. They coincide with
the singlet representation of the previous section since
the eigenvalues of $\k^i$ on the highest weight vectors coincide with
those of the singlet representation both in the unbarred and in the barred
case. $\diamondsuit$

It remains to realize the scalar product of ${\cal H}^{\vec{0}}$.
The relevance of the double realization
manifests itself in the

\begin{theorem}
The scalar product in ${\cal H}^{\vec{0}}$ can be realized in
configuration-space by
\be
<\phi_1|\phi_2>=\int d_qV ~~[\bar{\varphi}_1]^* \varphi_2,
\ee
where $\int d_qV$ is the integration first defined in \cite{fio2},
with a suitable normalization.
\end{theorem}
$Proof$.
Let $<\phi_1|\phi_2>':=RHS(7.55)$. Because of the lemma,
\be
<\phi_1|\phi_2>'=\int d_qV ~~e_{q^2}[-m^2x\cdot x]
[\varphi_1(q^{-1}x)]^* \varphi_2(x)~~~~~~~~~~m^2:=(1-q^2)q^{-2\rho_n-3}M^2;
\ee
$<~~,~~>'$ is a well defined inner product in ${\cal H}_{\vec{0}}$:
in fact, the integral (7.56) is well defined according to the definition
given in ref. \cite{fio2}, due to the presence of the
the $q^2$-gaussian damping factor
$e_{q^2}[-m^2x\cdot x]$, which can be taken as `` reference
function of the integration'' : in the frame of that definition
it suffices to expand
$[\varphi_1(q^{-1}x)]^* \varphi_2(x)$ in powers of $x$, perform the
integrations and resum the terms to obtain the above integral.
{}From the practical point of view it is convenient, however,
to choose directly $[\bar{\varphi}_0]^* \varphi_0$,
where $|\phi_0>=|\vec{0},\vec{0}>$ stands for the cyclic vector,
as (non-scalar) reference function of the integration (in the
sense mentioned in ref. \cite{fio2}), since, as we will
see below, all integrals of the form (165) can
be evaluated from this basic integral.

Because the integration satisfies Stoke's theorem with derivatives $\p,\bp$
\cite{fio2}, i.e. `` boundary terms '' vanish, and  formula (155) holds,
the complex conjugation $*$ followed by an exchange
of the two realizations acts as hermitean conjugation
of all differential operators of $u_q(e^N)$ with respect to the
scalar product of ${\cal H}^{\vec{0}}$. This ensures that
the inner product $<~~,~~>'$ preserves the orthogonality relations
between different vectors of the basis ${\cal B}_q$ of ${\cal H}^{\vec{0}}$
since they have different eigenvalues of the observables (115).

If $|\phi_i>={\cal D}_i|\phi_0>$,
${\cal D}_i\in u_q^{c,N}$ ($u_q^{c,N}$ was defined in formula (165)),
$i=1,2$, then because of formula (155)
\be
\int d_qV ~~[\bar{\varphi}_1]^* \varphi_2=
\int d_qV ~~[\bar{\varphi}_0]^* \rho({\cal D}_1^*{\cal D}_2|\phi_0>).
\ee
Only the $\rho$ image of the nonzero $|\phi_0>$-component of
${\cal D}_1^*{\cal D}_2|\phi_0>$, if any, will contribute to the above
integral. This  explains why the evaluation of all integrals of
the form (165) is reduced to the basic integral
$\int d_qV~[\bar{\varphi}_0]^* \varphi_0$, as claimed. Finally, we
normalize the integration so that $\int d_qV [\bar{\varphi}_0]^* \varphi_0=1$.
This concludes the proof that $<~~,~~>'=<~~,~~>$. $\diamondsuit$

{\bf Remark 1}
To recognize that the inner product $<~~,~~>'$ is sesquilinear it
is convenient to consider its
more symmetric but equivalent form
\be
<\phi_1|\phi_2>':=\int d_qV ~~\left([\bar{\varphi}_1]^* \varphi_2+
[\varphi_1]^* \bar{\varphi}_2\right)
\ee
(alternatively one could include in the RHS only the second term).
In fact, the sesquilinearity of the scalar product is immediate in this form.
That this form is equivalent to the former can be esaily understood.
Actually each one of the two terms separatly allows to formally realize the
$\dagger$-structure of equations (84) in terms of differential operators
acting on spaces of functions on $\rn_q^N$,
and the whole Hilbert space ${\cal H}^{\vec 0}$
is built up from a single cyclic
vector through application of elements of $u_q(e^N)$; as we have seen
all inner products
are therefore completely determined in terms respectively of the integrals
$\int d_qV ~~[\bar{\varphi}_0]^* \varphi_0$ and (167)
involving the cyclic state $|\phi_0>$. It sufficient to normalize both
equal to 1 to make them coincide.

{\bf Remark 2}. Formula (165) allows to explain from the point of
view of configuration space the regularizing effect of taking the
$q$-deformed version of the Euclidean Hopf-algebra instead of the classical
one. Only when $q\neq 1$ the damping factor
$e_{q^2}[-m^2x\cdot x]\neq 1$  in the integral (165)
makes the norms of eigevectors of the observables (115) finite.

{\bf Remark 3} The final lesson we learn from interpreting the results
of sections IV.1,IV.2 is the following. Requirement
$(114)_1$ of nonnegativity of the scalar product makes the
representations of $\hat u_q(e^N)$ of highest weight type.
If we stick for simplicity
to the highest weight singlet
representation of $u_q(e^N)$, then the fact that
power series in the coordinates $x^i$ (the ones belonging to subspaces
$\rho({\cal H}^{\vec{0}}),\br({\cal H}^{\vec{0}})\subset\FR$ which
we have constructed) arise as a natural basis to
identify elements of the singlet irrep of $u_q(e^N)$
(instead of the vectors obtained by applying combinations of
the elements of a Poincare'-Birkhoff-Bott basis of
$u_q^{c,N}$ to the highest weight state)
ultimately is traced back to the requirement $(114)_2$
that the scalar product is nondegenerate.
In fact, equation $(114)_2$ eliminates makes a huge amount of singular
vectors to appear within the space of
such combinations; having moded the latter out, we have shown that the
abovementioned power series in $\FR$ are sufficient to identify the
remaining vectors.

\subsection{Classical limit of the singlet irrep}

{}~~~In this section we just briefly sketch what we are allowed to mean
by `` $\lim_{q\rightarrow 1}\Gamma_q=\Gamma_{q=1}$ '', i.e. by saying that
an irrep of $u_q(e^N)$ goes to an Irrep $U(e^N)$ in the limit
$q\rightarrow 1$. If we compare the behaviour
of the representations $\Gamma_q$ of $u_q(e^N)$
under this limit with that of the representations
of $U_q^N$ (which is the `` compact '' subalgebra of $u_q(e^N)$), we find
important differences.
For simplicity we stick to
the singlet Irrep of $u_q(e^N)$;
the other Irreps of $u_q(e^N)$ can be obtained
by the tensor product(139), and the properties of the Irreps of $U_q^N$ are
essentially known.

The commuting observables
\be
p_0,(p\cdot p)_1,...,(p\cdot p)_{n-1},(p\cdot p)_n;
h_1,...,h_n~~~~~~~~~~~~~~~~~~(p_0\equiv 0~~~if~~N=2n+1).
\ee
 ($h_i:= ln_{q^2}(\k^i)$)
make up a complete set both when $q\neq 1$ and $q=1$. We have chosen
$h_i$ instead of $\k^i$ because it is the set of generators
$\{L^{i,j},h_i,p^i\}$ which has
classical commutation relations in the limit $q\rightarrow 1$.
The eigenvalues of the observables (115)
label the vectors of an orthonormal basis
${\cal B}_q$ (149) (of eigenvectors) of the singlet Irrep for
all $q\in \rn^+$; when $q=1$ the vectors of this basis are distributions
(see the configuration space realization of the preceding section),
i.e. ${\cal H}_{q=1}$ is no more a Hilbert space but the space of functionals
on some space of smooth functions on $\rn^n$, e.g. ${\cal S}(\rn^N)$.

For each fixed eigenvector $|\Pg,\J>$ the
eigenvalues $j_i$ of $h_i:=log_{q^2}(\k^i)$ don't depend on $q$ and
are integers; whereas the eigenvalues
of $(p\cdot p)_i$ (non-uniformly) `` collapse '' to $M^2$:
\be
\lim_{q\rightarrow 1}c_i(q,\Pg)=M^2
{}~~~~~~~~~~where~~~~~(p\cdot p)_i|\Pg,\J>=:c_i(q,\Pg)|\Pg,\J>.
\ee
Therefore all vectors $|\Pg,\J>$ of ${\cal B}_q$ with fixed $\J$
would have separatly
{}~the same limit when $q\rightarrow 1$, and the latter would
coincide with a vector~ of~ ${\cal B}_{q=1}$;~~ consequently,
$\{\lim_{q\rightarrow 1}|\Pg,\J>~,~|\Pg,\J>\in {\cal B}_q\}\neq {\cal
B}_{q=1}$.
Therefore the limit
$\lim_{q\rightarrow 1}\Gamma_q=\Gamma_{q=1}$ cannot be given a literal sense.

However, we can give a weaker sense to the above limit, as we are going to
explain.
Assume that $q=1$ and the vector
$\vec{r}:=(r_h,....,r_{n-1},r_n)$ consists of components
$r_n\in \rn^+$, $0\le r_i\le 1$  $h\le i\le n-1$, $\J\in \zn^n$; let
$||\vec{r},-\J>\in {\cal B}_{q=1}$
be the vector with eigenvalues $h_i=-j_i$, $(p\cdot p)_i=(p\cdot p)_{i+1}r_i$,
$(p\cdot p)_n=M^2r_n$.  Define functions
$\tilde {\pi}_i(r_i,q):=\left[ ln_{q^2}(r_i)\right]$ ($[a]$ denotes the
integral
part of the number $a\in \rn$), and set
$\tilde{\Pi}_i:=\sum\limits_{l=i}^{n-1}\tilde {\pi_l}$.
Note that
$lim_{q\rightarrow 1^{\pm}}\tilde{\pi}_i=\mp \infty$
whenever $r_i<1$.
If $q<1$ (as we have
assumed in all this chapter) and $(1-q)$ is
sufficiently small then $-j_i\le \tilde{\pi}_{i-1}$, and we can define
$$
|\psi_{\vec{r},-\J}>:=|\Pg=\tilde{\vec{\pi}}(\vec{r},q),-\J>
\propto
\La^{\tilde {\pi}_n}(p^n)^{\tilde {\pi}_{n-1}}...(p^{h+1})^{\tilde {\pi}_h}
$$
\be
\cdot (L^{-n,n-1})^{J_n+\tilde{\Pi}_{n-1}}...(L^{-2,1})^{J_2+\tilde{\Pi}_1}
\cases{(L^{-1,0})^{J_1+\tilde{\Pi}_0}|\vec{0},\vec{0}>~~~~~if~~N=2n+1 \cr
(p^{- sign(J_1+\tilde{\Pi}_1)\cdot 1})^{|J_1|+\tilde{\Pi}_1}|\vec{0},\vec{0}>
{}~~~~~if~~N=2n, \cr}
\ee
where $J_i$ are related to $j_i$ by formula (151).
The weak sense which can be given to the limit
$\lim_{q\rightarrow 1}\Gamma_q=\Gamma_{q=1}$ is at least that
for any $||\vec{r},-\J>\in {\cal B}_{q=1}$ and small $\ve>0$
we can find a $q<1$ and a vector
$|\psi_{\vec{r},-\J}>\in {\cal B}_q\subset {\cal H}_q$ such that
the corresponding eigenvalues of the observables $(p \cdot p)_i$ differ
by less than $\ve$: indeed, we only need to set
$q\equiv 1-\ve q^{-r_n}$, solve for $q$ and define $|\psi_{\vec{r},-\J}>$
as in formula (170). This expresses in a precise form the fact that,
roughly speaking, the tori $(p\cdot p)_i=c_i(\Pg)$, $\Pg\in\nn^{n-h}\times\zn$,
get `` dense '' in all the momentum space when $q\rightarrow 1$.

Note that the convergence of the q-deformed eigenvalues
(selected in this way) to the classical ones is uniform
only   `` on the states localized on the sphere $(p\cdot p)_n=c$ '',
i.e. within each subspace characterized by a fixed value $(p\cdot p)_n=c$.

In the limit $q\rightarrow 1$ the `` parity asymmetry '' in the spectrum
of the observables (115) noticed at the end of section IV.1.4 disappears,
in the sense that the range of each $j_i$ (as a function of the
square momenta) becomes the whole set $\zn$,
whenever $r_{i-1}<1$, i.e. $(p\cdot p)_{i-1}<(p\cdot p)_i$, i.e.
`` almost everywhere '' in momentum space.
(In fact, the condition $(p\cdot p)_{i-1}=(p\cdot p)_i$
fixes a cylinder in the classical momentum space $\rn^N_{\vec{p}}$,
this is a subset of $\rn^N_{\vec{p}}$ of zero measure).
The same is true also in the Irreps with highest weight $\neq 0$.


\section{Appendix}

\subsection{Proof~of~Proposition~3}
Since $U_q^N\ap U_{q^{-1}}(so(N))$ and $Fun(SO_{q^{-1}}(N))$
are known \cite{frt} to be Hopf dual, we only need to check the compatibility
of the new commutation relations of either
algebra with the coalgebra structure of its dual partner.

The pairing of the homogenous Hopf sub-algebras on the generators
$L^{\pm}$ \cite{frt} takes the form
\be
<L^{\pm,i}_{~~h},T^j_k>=R_{q^{-1}~hk}^{\pm 1~ij}~~~~~~~~~~~~~~R_{q^{-1}}^+=
\hat R_{q^{-1}} P,~~~~~
R_{q^{-1}}^{-1}=\hat R_{q^{-1}}^{-1} P;
\ee
here one uses the basis
$\{L^{\pm,i}_{~~j}\}$ of the subalgebra
$Fun(SO_{q^{-1}}(N))^*_{reg}\subset U_{-h}so(N)$, in the notation of Ref
\cite{frt}. Using the transformation \cite{frt} from these generators
to the Drinfeld-Jimbo ones, and the transformation \cite{fio4} from the
latter to our generators $\{\M,\k\}$, one can easily arrive at some
useful pairing relation between the $T's$ and the $\M,\k$:
\be
<\k^i,T^h_k>=\delta^h_kq^{2(\delta^{-i}_k-\delta^i_k)}
\ee
$$
<\M^{-m,m-1},T^j_h>=\f{q^{\rho_m+1}}{1-q^{-2}}\hat R^{-1~1-m,j}_{q^{-1}~h,-m}
\cdot q^{\delta^m_k-\delta^{-m}_k}
$$
\be
<\M^{1-m,m},T^j_h>=\f{q^{\rho_m-1}}{1-q^2}\hat R^{~~~j,-m}_{q^{-1}~1-m,h}
\cdot q^{\delta^{m-1}_h-\delta^{1-m}_h}.
\ee
Note that the matching of conventions of Ref. \cite{frt} with ours
requires: 1) shifting the $i$ indices in \cite{frt}, all indices
by $-n-1$ in the
$N=2n+1$ case, the first $n$ indices by $-n-1$ and the remaining by $-n$ in
the $N=2n$ case; 2) the inverse ordering of the spots of the $so(N)$
Dynkin diagrams.

Let us consider the unbarred Hopf algebras. As a first step we consider the
compatibility between the
algebraic relations within $U_q(e^N)$ and the coalgebra structure
of $Fun(E_{q^{-1}}^N)$, starting from relation $(69)_1$:
\be
<\La\p_n-q^{-1}\p_n\La,\cases{w \cr T^i_j\cr}>=
<\La\ot\p_n-q^{-1}\p_n\ot\La,\cases{w\ot w\cr T^i_h\ot T^h_j\cr}>=0
\ee
trivially, whereas
\be
<\La\p_n-q^{-1}\p_n\La,y^i>=
<\La\ot\p_n-q^{-1}\p_n\ot\La,y^i\ot 1+w^{-1}T^i_h\ot y^h>=
(<\La,w^{-1}>-q^{-1})\delta^i_n\stackrel{(89)_1}{=}0;
\ee
hence one easily realizes that the check with higher powers in $w,T,y$
in the RHS is trivial, in other words
$<\La\p_n-q^{-1}\p_n\La,Fun(E_{q^{-1}})>$=0.
Now we go on by writing down explicitly only the nontrivial checks. The
check of relations $(69)_2,(69)_3$ is trivial, the check of relations
$(69)_4$ is trivial
when these relations are paired with $w^{\pm 1},T^i_j$, whereas:
$$
<\k^i\p_n-q^{2\delta^i_n}\p_n\k^i,y^j>=
<\k^i,T^j_l><\p_n,y^l>-q^{2\delta^i_n}<\p_n,y^j><\k^i,1>
$$
\be
\stackrel{(172)}{\propto}
\delta^j_n(q^{2\delta^i_n}-q^{2\delta^i_n})=0~~~~~~~~~~~~i>0;
\ee
\be
<[\M^{-i,i-1},\p_n]_{a_{i,-n}},y^j>=<\M^{-i,i-1},T^j_l><\p_n,y^l>
\stackrel{(173)}{\propto}\hat R^{-1~1-i,j}_{q^{-1}~n,-i}=0,
\ee
\be
<[\M^{1-m,m},\p_n],y^j>=<\M^{1-m,m},T^j_l><\p_n,y^l>
\stackrel{(173)}{\propto}\hat R^{~~~j,-m}_{q^{-1}~1-m,n}=0
,~~~~~~~n>m>h
\ee
\be
<\left[\M^{1-n,n},[\M^{1-n,n},\p_n]_q\right]_{q^{-1}},\y^j>=
<(\M^{1-n,n})^2,T^j_l><\p_n,y^l>\stackrel{(173)}{\propto}
\hat R^{~~~j,-n}_{q^{-1}~1-n,a}\hat R^{~~~a,-n}_{q^{-1}~1-n,n}=0;
\ee
Here we have used the explicit form of $\hat R_{q^{-1}}$ \cite{frt,car}.
The  second step is to consider the compatibility between
the algebraic relations in $Fun(E_{q^{-1}}^N)$ and the coalgebra
structure of $U_q(e^N)$. The nontrivial checks follow:
\be
<\p_n,wy^j-qy^jw>=(<\La,w>-q)\delta_n^j=\stackrel{(89)_1}{=}0,
\ee
$$
<\p_n,y^iT^j_k-\hat R^{-1~ij}_{q^{-1}~lm}T^l_ky^m>=
<\p_n,y^i>\delta^j_k-<(\k^n)^{\f 12},T^l_k>\hat R^{-1~ij}_{q^{-1}~ln}
$$
\be
\stackrel{(172)}{=}\delta^i_n\delta^j_k- \hat R^{-1~ij}_{q^{-1}~kn}
q^{\delta_n^k-\delta^{-k}_n}=0,
\ee
\be
<\p_n\p_n,{\cal P}_{A~hk}^{~~ij}y^hy^k>\stackrel{(89)}{\propto}
{\cal P}_{A~hk}^{~~nn}=0.
\ee

The proof for the barred Hopf algebras is analogous. $\diamondsuit$

\subsection{Proof of Theorem 5}

We will denote by
${\cal H}_{\vec{0},\vec{\mu}}\subset {\cal H}_{\vec{0}}$
the eigenspace of all $\k^i$'s with eigenvalues  $\mu_i$.
We can easily show that ${\cal H}_{\vec{0},\vec{\mu}}$ is finite
dimensional $\forall \vec{\mu}\in {\bf R}^n$.
In fact, if $|\phi>\in{\cal H}_{\vec{0},\vec{\mu}}$, $|\phi>\neq 0$,
the whole ${\cal H}_{\vec{0},\vec{\mu}}$ has to be obtained by
applying zero-graded (w.r.t. the grading of $\k^i$) operators
obtained as polynomials of elements $L_{+a},L_{-a}$ ($a$ are the roots
of $U_q^N$) of the Poincare'-Birkhoff-Witt
bases of the subalgebras $u_q^{+,N},u_q^{-,N}$ introduced in section III.3;
But these operators make up a finite-dimensional vector space ${\cal L}_0$
within each irrep. Indeed,
note that the $(n+1-h)$ casimirs of $\hat u_q^N$ belong to ${\cal L}_0$;
they are proportional to the identity operator within each irrep.
Then, one can easily realize that within each irrep
there is only a finite number of linearly independent polynomial
operators of the kind mentioned, since one can iteratively
mode out casimirs from polynomials of higher degrees in the
roots, so as to
reduce them to polynomials of lower degree, until one reduces them to
combinations of the elements of a finite `` basis ''
of low degree polynomials.

Assume for brevity that all eigenvalues $\mu_i$ are positive.
Let $L$ be one of the positive roots, $L^-$ its Cartan-Weyl partner;
then their
commutation relation is of the form
\be
[L,L^-]_a=c \f{(1+d)-\k}{1-a^{-1}},~~~~~~~~~0<a<1,c>0~~~~~~~~~~~~~~~~
[\k,L]_{a^{-2}}=0=[\k,L^-]_{a^2},
\ee
where $k$ is an element of the Cartan subalgebra, and in the sequel

$a=\cases{q~~~~if~~~N=2n+1~~and~~L=L^{0l}\cr q^2~~~~otherwise\cr}$.

\begin{lemma}
Let $L,L^-,k$ satisfy relations
(183) and let $L^-L$ be hermitean positive definite
(both conditions are
satisfied if $L$ is a simple roots, for instance).
For any $\vec{\mu}$ there $\exists m\in \nn$ such that
$(L)^m{\cal H}_{\vec{0},\vec{\mu}}={\bf 0}$; if $1+d<0$,
then $\exists m'\in \nn$ such that
$(L^-)^{m'}{\cal H}_{\vec{0},\vec{\mu}}={\bf 0}$
\end{lemma}

$Proof$. Let $\{|f_i>\}_{i\in I}$ be a basis of
${\cal H}_{\vec{0},\vec{\mu}}$ consisting of eigenvectors of $L^-L$
and let $l_i$ be the corresponding eigenvalues. By recursively applying
commutation relation (7.62) we find that $L^m|f_i>$, $m\in \nn$,
is an eigenvector of the positive definite operator $L^-L$:
$$
(L^-L)(L^m)|f_i>=l(l_i,m)(L^m)|f_i>
$$
\be
l(l_i,m):=\left[l_ia^{-m-1}-
(m+1)_{a^{-1}}ca^{-1}\f{(1+d)-\lambda a^{-m}}{1-a^{-1}}\right].
\ee
where
$\lambda$ denotes the eigenvalue of $\k$ in ${\cal H}_{\vec{0},\vec{\mu}}$.
We see that $l(l_i,m)\le l(b,m)$ ($b\ge l_i$ $\forall i$),
and that $l(b,m)$ would get negative for large m, unless there
exists a $m \in \nn$ such that $L^m|f_i>=0$ $\forall i\in I$.
Similarly one proves the second part of the Lemma. $\diamondsuit$.

Applying repeatedly Serre relations (50)-(52) one can prove the
following
\begin{lemma}
Let $|j|<k$, $k\ge i\ge h+1$.
\be
[\M^{1-i,i},\M^{j,k}]_a=0~~~~~~~~~~~~a=\cases{q~~~~~~if~~~i=k,~~j\neq
1-i,i-1\cr
q^{-1}~~~~~~if~~~i<k,~~j=i,1-i \cr 1~~~~~~~if~~~i<k,~~j\neq \pm i,\pm(i-1);\cr}
\ee
as a consequence, if $k>j>i\ge h$
\be
[\M^{j,k},\M^{i,k}]_{q^{-1}}=0~~~~~~~~~~~~~~~~~~~~
[\M^{-j,k},\M^{-i,k}]_q=0.
\ee
The same holds if we replace the $\M$ by the $L$ roots.
\end{lemma}

$Proof~of~theorem~5$.
Because of proposition 8, it is sufficient to prove the theorem
within ${\cal H}_{\vec{0}}$. We prove only the first part of the thesis,
stating the existence of a highest weight vector and its uniqueness.
The rest of the
proof will be given elsewhere \cite{fio7},
where we will use
some explicit knowledge about the casimirs of $\hat u_q(e^N)$.

Given any $|\psi_0>\in{\cal H}_{\vec{0}}$, let $k^i|\psi_0>=\lambda_i|\psi_0>$
and apply lemma 9 to
${\cal H}_{\vec{0},\vec{\lambda}}$ by
setting $L\equiv L^{01}$
if $N=2n+1$ and $L\equiv L^{\pm 1,2}$ if $N=2n$; we will respectively
determine integers $p,p_{\pm}\in\nn$ such that
$$
(L^{01})^{p+1}|\psi_0>=0~~~~~~~~~
(L^{01})^p|\psi_0>\neq 0~~~~~~~~~~~~~~~if~~~N=2n+1
$$
\be
\cases{(L^{1,2})^{p_++1}(L^{-1,2})^{p_-}|\psi_0>=0=
(L^{1,2})^{p_+}(L^{-1,2})^{p_-+1}|\psi_0>,\cr
(L^{1,2})^{p_+}(L^{-1,2})^{p_-}|\psi_0>\neq 0 \cr}~~~~~~~~~~~if~~~N=2n.
\ee
If we define
\be
|\psi_1>:=\cases{(L^{01})^p|\psi_0>~~~~~~~~~~~~~~~if~~~N=2n+1 \cr
(L^{1,2})^{p_+}(L^{-1,2})^{p_-}|\psi_0>~~~~~~~~~~~~~~~~if~~~N=2n, \cr}
\ee
this means that $|\psi_1>$
is annihilated by $U_q^{+,3}$ (resp. $U_q^{+,4}$).

Now the proof goes on by induction. Assume that we have determined a
nontrivial vector $|\psi_{j-1}>\in{\cal H}_{\vec{0}}$ such that
$U_q^{+,2j-1+h}|\psi_{j-1}>=\{{\bf 0}\}$, $j=2,...,n$.
Moreover, assume that we have determined
integers $p_l\in \nn$, $l=j-1,j-2,...,i$, $j-1\ge i\ge 2-j$, such that
\be
\cases{
|\psi_{i,j}>:=(L^{i,j})^{p_i}(L^{i+1,j})^{p_{i+1}}...(L^{j-1,j})^{p_{j-1}}
|\psi_{j-1}>\neq 0,\cr
L^{i,j}|\psi_{i,j}>=0~~~~~~~~~~U_q^{+,2j-1+h}|\psi_{i,j}>=0\cr}
\ee
Then we can determine an integer $p_{i-1}\in \nn$ such that
\be
\cases{
|\psi_{i-1,j}>:=(L^{i-1,j})^{p_{i-1}}(L^{i,j})^{p_i}...(L^{j-1,j})^{p_{j-1}}
|\psi_{j-1}>\neq 0,\cr
L^{i-1,j}|\psi_{i-1,j}>=0~~~~~~~~~~U_q^{+,2j-1+h}|\psi_{i-1,j}>=0\cr}
\ee
In fact, on one hand we set $L\equiv L^{i-1,j}$ and try to apply lemma 3.
It is well known that when $L$ is not simple the Cartan-Weyl partner of
$L$, $L^-$, differs
from $(L^{i-1,j})^{\dagger}\propto L^{-j,1-i}$; its polynomial expression in
terms of simple negative roots can be obtained from the one
of $L^{-j,-i}$ by the replacement $q\rightarrow q^{-1}$. Viceversa,
the polynomial expression in terms of simple positive roots of
$L':=(L^-)^{\dagger}$ can be obtained from the one
of $L^{i-1,j}$ by the same replacement. For instance,
\be
L:=L^{-1,3}\propto [L^{-1,2},L^{-2,3}]_q~~~~~~~
(L)^{\dagger}=L^{-3,1}\propto [L^{-3,2},L^{-2,1}]_q;
\ee
\be
L^-\propto [L^{-3,2},L^{-2,1}]_{q^{-1}}~~~~~~~
L'\propto [L^{-1,2},L^{-2,3}]_{q^{-1}}.
\ee
One can easily verify that in any case we can find
$u'\in U_q^{+,N-2},~~~u\in u_q^{+,N}~~a>0$
such that
\be
L'=aL+uu'~~~~~~~\Rightarrow~~~~~~~~~L^-L'=aL^-L+ L^-uu'.
\ee
Now the second term in the RHS can be neglected because it
always gives zero when applied to $(L^{i-1,j})^m|\psi_{i,j}>$,
because of the induction hypothesis (189).
Therefore we are in the conditions
to apply lemma 3 to the subspace ${\cal H}_{\vec{0},\lambda}$
containing $|\psi_{i,j}>$,
since $L^-L'$ is positive definite and $L,L^-$ belong to
a Cartan-Weyl triple.

On the other hand, one can easily show that the
generators $L^{1-k,k}$ ($k=h+1,h+2,...,j-1$) of $u_q^{+,2j-1+h}$
 annihilate $|\psi_{i-1,j}>$ as well,
by application of the commutation relations (185),(186), their consequences
\be
L^{1-k,k}(L^{-k,j})^p=q^p(L^{-k,j})^pL^{1-k,k}+aq^p(L^{-k,j})^{p-1}
L^{1-k,j}~~~~~~~~~~~a>0
\ee
\be
L^{1-k,k}(L^{k-1,j})^p=q^p(L^{-k,j})^pL^{1-k,k}+a'q^p
(L^{-k,j})^{p-1}L^{k,j}~~~~~~~~~~~~a'>0
\ee
and the induction hypothesis. In the least simple case, $i-1\le -k$,
for instance,
$$
L^{1-k,k}|\psi_{i-1,j}>\stackrel{(185)_c}{\propto}...L^{1-k,k}
(L^{-k,j})^{p_{-k}}...(L^{j-1,j})^{p_{j-1}}|\psi_{j-1}>
$$
$$
\stackrel{(194)}{\propto}
...[(L^{-k,j})^{p_{-k}}L^{1-k,k}+a(L^{-k,j})^{p-1}
L^{1-k,j}](L^{1-k,j})^{p_{1-k}}...(L^{j-1,j})^{p_{j-1}}|\psi_{j-1}>
$$
$$
\stackrel{(189)_b}{=}
...L^{1-k,k}(L^{k-1,j})^{p_{k-1}}...(L^{j-1,j})^{p_{j-1}}|\psi_{j-1}>
$$
$$
\stackrel{(195)}{\propto}...[(L^{k-1,j})^{p_{k-1}}L^{1-k,k}+a'
(L^{k-1,j})^{p_{k-1}-1}
L^{k,j}](L^{k,j})^{p_k}...(L^{j-1,j})^{p_{j-1}}|\psi_{j-1}>
$$
\be
\stackrel{(189)_b}{\propto}......(L^{j-1,j})^{p_{j-1}}L^{1-k,k}|\psi_{j-1}>
\stackrel{(189)_c}{=} 0.
\ee
The dots stand for the same powers of the roots $L$ which appear in the
definition $(189)_a$ of $|\psi_{i-1,j}>$.

Finally let us define
\be
|\psi_j>:=
(L^{1-j,j})^{p_{1-j}}(L^{2-j,j})^{p_{2-j}}
...(L^{j-1,j})^{p_{j-1}}|\psi_{j-1}>;
\ee
it is easy to show the relation $U_q^{+,2j+1+h}|\psi_j>=\{{\bf 0}\}$,
i.e. the induction hypothesis for the subsequent step.

In fact, on one hand
$L^{1-j,j}|\psi_j>=0$ trivially, because of equation (190) and the definition
(197); on the other, $U_q^{+,2j-1+h}|\psi_j>=\{{\bf 0}\}$ because
$U_q^{+,2j+1+h}$ annihilates by construction each of the vectors $(189)_a$

The first part of the thesis is proved if we set $|\phi>=|\psi_{1-n,n}>$.

To prove the uniqueness of the highest weight vector in the case
$N=2n+1$ (resp. the $\zn$-structure of ${\cal H}_G$ in the case $N=2n$)
we can proceed as follows. Let $|\phi>,|\phi'>$
the highest weight vectors constructed by the above procedure starting
from two vectors $|\psi_0>,|\psi'_o>\in {\cal H}_{\vec{0}}$. The
$n-h$ casimirs of $\hat u_q(e^N)$ different from $(p\cdot p)_n$ are
independent functions of the generators and when acting on highest weight
vectors as $|\phi>,|\phi'>$ will reduce to $n-h$ independent functions
of $\k^i$ only. In fact, the casimirs are sum of zero-graded (w.r.t. $\k^l$)
monomials in $L,\k$ which can be written in a form where
$L\in u_q^{+,N}$ stand at the right of $L\in u_q^{-,N}$, and
the former annihilate $|\phi>,|\phi'>$. But since
these casimirs take the same values on ${\cal H}_{\vec{0}}$, this implies
that $|\phi>,|\phi'>$ have the same $\k^i$ eigenvalues $\forall i$ if $N=2n+1$,
$\forall i>1$ if $N=2n$ (in fact, $(p^{\pm 1})^r|\phi>$ are
highest weight eigenvectors $\forall r\in \nn$). This implies the
claim on the dimensionality of ${\cal H}_G$,
since as a consequence
${\cal O}|\phi>=|\phi'>$ implies that ${\cal O}$ is a zero-graded
(w.r.t. to $\k^i$, $i=h+1,h+2,...,n$) operator, and up to a
constant such an operator acts on
a highest weight vector as the identity,
in the case $N=2n+1$, and as some powers of $p^{\pm 1}$ in
the case $N=2n$. $\diamondsuit$

\subsection{Proof~of~theorem~6}

It is immediate to check that commutation relations
(147),(148) hold when applied to $|\vec{0},\vec{0}>$. Let
${\cal H}_{\vec{0},(\J),\le}:=Span_{\cn}\{|\vec{0},\vec{l},...>~
|~l^i\ge j^i\}$.
The proof of these equations is by induction in the
weights $\J$, i.e. assuming that they hold within
${\cal H}_{\vec{0},(\J),\le}$ we show that they hold
when applied on ${\cal H}_{\vec{0},\J-\e_l+e_{l-1},\le}$.

As a consequence of formulae (147),(148) one can easily show that within
${\cal H}_{\vec{0},(\J),\le}$
\be
L^{-i,i-1}L^{1-i,i}=
\cases{q^{-\f
32}\f{(\k^1)^{-1}-1}{(1-q)(q^{-1}-q)}~~~~~if~~~N=2n+1~~and~~i=1\cr
\f{q^{-2}\k^1(1-(\k^2)^{-1})}{(1-q^2)^2}~~~~~if~~~N=2n~~and~~i=2 \cr
q^{2\rho_i}\f{[(\k^i)^{-1}-1][1-q^{-2}\k^{i-1}]}{(q-q^{-1})^2}~~~~otherwise.
\cr}
\ee

First part of the thesis: if $|\chi>$ is singular, then for any
$u\in \hat u_q^{-,N}$
$u|\chi>$ is. It is sufficient to consider only $u=p^{\pm 1}$
or $u=L^{-l,l-1}$,
and vectors $|\chi>$ of the form
$|\chi>=vg_iv'|0>$, where $v,v'$ are two ordered polynomials in the generators
of $\hat u_q^{-,N}$; in fact, by the induction hypothesis the
latter are singular. Any other singular vector can be expressed as a
combination
of singular vectors of this form.

We consider first relations (147) in the case $N=2n$.
Since $[g_1,u_q^{-,N}]=0$,
$|\chi>$ is of the form $|\chi>=vg_1|\vec{0},\vec{0}>$ $v\in \hat u_q^{-,N}$.
The vector $|\chi'>:=L^{-2,1}|\chi>$ is trivially orthogonal to any
vector corresponding to a different weight. In addition it is orthogonal
to the only two independent vectors $|1>,|2>$ with the same weight, which
are (according to formula (149) of the induction hypothesis)
\be
|1'>:=L^{-2,1}|1>:=L^{-2,1}vL^{-2,-1}|\vec{0},\vec{0}>~~~~~~~~~~~~
|2'>:=L^{-2,1}|2>:=\f{L^{-2,1}vL^{-2,1}
q^{-2}p^{-1}p^{-1}}{(1-q^2)(p\cdot p)_1}|\vec{0},\vec{0}>.
\ee
In fact, one can easily check that for any $v$ as considered above,

$(v^*L^{-1,2})L^{-2,1}=a~L^{-2,1}(v^*L^{-1,2})+\k v^*$ ($\k$ being
an element of the Cartan subalgebra), $[v^*,L^{-12}]=0$, implying
$$
<1'|\chi'>\propto <\vec{0},\vec{0}|L^{1,2}(v^*L^{-1,2})L^{-2,1}|\chi> \propto
<\vec{0},\vec{0}|L^{1,2}L^{-2,1}(v^*L^{-1,2})|\chi> + b
<\vec{0},\vec{0}|L^{1,2}v^*|\chi>
$$
\be
=<\vec{0},\vec{0}|L^{-2,1}L^{1,2}|\chi>+ c<2||\chi> + b'<1|\chi>=0;
\ee
in the last identity we have used the singularity of $|\chi>$ and the
fact that $<\vec{0},\vec{0}|L^{-2,1}=0$. Similarly one shows that
$<2'|\chi'>=0$. Summing up, $|\chi'>$ is singular. The corresponding result
for $\tilde g_1$ is a direct consequence of the above.

Now we consider the remaing commutation relations (148) (for even and
odd $N$ at the same time). One can easily check that they imply the
relations
\be
L^{-m,m-1}g_i=g_if_lL^{-m,m-1}~~~~~~~~~~~~f_m:=\cases{
q\f{1-q^{-2}\k^i}{1-\k^i}~~~if~~m=i,i+1 \cr
q\f{1-q^{-2}\k^{i+1}}{1-\k^{i+1}}~~~if~~m=i+2 \cr
q^{-1}\f{1-\k^{i-1}}{1-q^{-2}\k^{i-1}}~~~if~~m=i-1 \cr
1~~~otherwise; \cr}
\ee
since they hold in ${\cal H}_{\vec{0},(\J),<}$, we can reduce any singular
vector to one of the form $|\chi>=g_iu''|0>$. Using commutation relations
(148) and  relations (149) only in ${\cal H}_{\vec{0},(\J),<}$
one can easily show that
\be
L^{-l,l-1}L^{1-l,l}g_i=s_{i,l}(\k) g_i;~~~~~~~~~~~or~~equivalently:
{}~~~~~~~~~~L^{1-l,l}L^{-l,l-1}g_i=s'_{i,l}(\k) g_i.
\ee
The vector $|\chi'>:=L^{-l,l-1}|\chi>$ is trivially orthogonal to any
vector corresponding to a different weight; in addition it has zero norm
\be
<\chi'|\chi'>\propto <\chi|s'_{i,l}|\chi>=0
\ee
because of the preceding relation and the fact that $|\chi>$ is singular;
therefore $|\chi'>$ is singular itself.

{}~~

Second part of the thesis: if $|\phi>\in {\cal H}_{\vec{0},(\J),\le}$ is
nonsingular, then $|\phi'>:=g_i|\phi>$ is singular.

Again, we consider first relations (147)
in the case $N=2n$. As already noticed,
it is easy to write this vector in the form $|\phi'>=vg'_1|\vec{0},\vec{0}>$;
then the proof that $|\phi'>$ is singular goes as for $|\chi'>$.

As for the remaining relations, using formulae (132),(130)
we find, when $i> h+1$
$$
[L^{-i,i+1}L^{1-i,i},g_i]_{q^2}=q^{2\rho_{i+1}}\k^i\f{1-q^2(\k^{i+1})^{-1}}
{1-q^{-2}\k^i}L^{-i,i-1}L^{1-i,i}+
$$
\be
q^{2\rho_i}\f{1-q^{-2}(\k^{i-1})^{-1}}
{1-q^{-2}\k^i}\left(q^2L^{-i-1,i}L^{-i,i+1}+q^{2\rho_{i+1}}
\f{1-\k^i(\k^{i+1})^{-1}}{1-q^{-2}}\right);
\ee
within ${\cal H}_{\vec{0},(\J),\le}$ we can replace the operators
$L^{-i,i-1}L^{1-i,i}$, $L^{-i-1,i}L^{-i,i+1}$ in the RHS by their expressions
(202), and we find that the RHS vanishes. A similar argument can be
used when  $i=h+1$, and also when the order of the two positive roots
appearing in the LHS is reversed. We find that on
${\cal H}_{\vec{0},(\J),\le}$
\be
[L^{-i,i+1}L^{1-i,i},g_i]_a=0~~~~~~~~
[L^{1-i,i}L^{-i,i+1},g_i]_a=0~~~~~~~~~~~~~~~a=\cases{q~~~~if~~N=2n+1,~~i=1 \cr
q^2~~~~otherwise.\cr}
\ee
Now consider the vector $g_i|\vec{0},-\J>$. On one hand,
 it is trivially orthogonal to
any other vector of ${\cal H}_{\vec{0}}$ of different weight; on the other, it
is orthogonal both to $|1>:=L^{-i-1,i}L^{-i,i-1}|\vec{0},-\J>$ and
$|2>:=L^{-i,i-1}L^{-i-1,i}|\vec{0},-\J>$
$$
<2|g_i|\vec{0},-\J>\propto <\vec{0},-\J|g_i L^{-i,i+1}L^{1-i,i}|\vec{0},-\J>=0
$$
\be
<1|g_i|\vec{0},-\J>\propto <\vec{0},-\J|g_i L^{1-i,i}L^{-i,i+1}|\vec{0},-\J>=0,
\ee
where we have used equations (204), the fact that the vectors
$g_i L^{-i,i+1}L^{1-i,i}|\vec{0},-\J>$,

$g_i L^{1-i,i}L^{-i,i+1}|\vec{0},-\J>$
belong to ${\cal H}_{\vec{0},\J,\le}$ and the induction hypothesis.
Therefore the vector $g_i|\vec{0},-\J>$ is singular in ${\cal H}_{\vec{0}}$ and
must be set equal to zero. $\diamondsuit$

{}~~

\section*{Acknowledgments}

I thank L. Bonora for encouragement and fruitful discussions.

{}~~

\end{document}